\documentclass[12pt]{article}
\usepackage{nicefrac}
\usepackage{amsfonts}
\usepackage{amsmath}
\usepackage{amsthm}

\advance \topmargin by -\headheight
\advance \topmargin by -\headsep
     
\evensidemargin \oddsidemargin
\def\rf#1{(\ref{eq:#1})}
\def\lab#1{\label{eq:#1}}
\def\nonu{\nonumber}
\def\br{\begin{eqnarray}}
\def\er{\end{eqnarray}}
\def\be{\begin{equation}}
\def\ee{\end{equation}}

\def\ba{\be\begin{array}{c}}
\def\ea{\end{array}\ee}

\def\lb{\lbrack}
\def\rb{\rbrack}

\def\({\left(}
\def\){\right)}
\def\v{\vert}                     %% vertical bars

\def\bc{\begin{center}}
\def\ec{\end{center}}
\def\Tr{\mathop{\rm Tr}}                  % Tr - big trace
\def\vp{{\varphi}}
\newcommand{\dder}[2]{\frac{d{#1}}{d{#2}}}
\newcommand{\pder}[2]{\frac{\partial{#1}}{\partial{#2}}}
\def\a{\alpha}
\def\b{\beta}

\def\d{\delta}

\def\eps{\epsilon}

\def\g{\gamma}

\newcommand{\h}{\frac{1}{2}}
\def\l{\lambda}

\def\p{\phi}

\def\pa{\partial}
\def\pr{\prime}

\def\s{\sigma}

\def\cA{{\cal A}}

\def\cD{{\cal D}}

\def\cF{{\cal F}}
\def\cG{{\cal G}}
\def\cH{{\cal H}}

\def\cJ{{\cal J}}
\def\cK{{\cal K}}
\def\cL{{\cal L}}
\def\cM{{\cal M}}

\newcommand{\ct}[1]{\cite{#1}}
\newcommand{\bi}[1]{\bibitem{#1}}
\def\cgh{{\widehat {\cal G}}}
\numberwithin{equation}{section}

\begin{document}

%\today

\begin{center}
{\large\bf Supersymmetry and the KdV equations   }
\end{center}
\begin{center}
{\large\bf for Integrable Hierarchies   }
\end{center}
\begin{center}
{\large\bf with a Half-integer Gradation}
\end{center}
\normalsize
\vskip .4in

\begin{center}
 H. Aratyn

\par \vskip .1in \noindent
Department of Physics \\
University of Illinois at Chicago\\
845 W. Taylor St.\\
Chicago, Illinois 60607-7059\\
\par \vskip .3in

\end{center}

\begin{center}
J.F. Gomes
and A.H. Zimerman

\par \vskip .1in \noindent
Instituto de F\'{\i}sica Te\'{o}rica-UNESP\\
Rua Pamplona 145\\
01405-900 S\~{a}o Paulo, Brazil
\par \vskip .3in

\end{center}

\begin{abstract}
Supersymmetry is formulated for integrable models 
based on the $sl(2|1)$ 
loop algebra endowed with a principal gradation.
The symmetry transformations which have half-integer grades
generate supersymmetry.
The $sl(2|1)$ loop algebra leads to $N=2$ supersymmetric mKdV and 
sinh-Gordon equations.
The corresponding $N=1$ mKdV and sinh-Gordon equations 
are obtained via reduction induced by twisted 
automorphism.
Our method allows for a description of a non-local symmetry 
structure of supersymmetric integrable models.

\end{abstract}

\section{\sf Introduction}
\label{section:intro}
We consider a class of models based on the superalgebras
$osp(1|2)$ and $sl(2|1)$ with the principal gradation.
We show that supersymmetry, when present, appears naturally as a part of the 
graded symmetry structure of integrable models based on 
superalgebras.
Our formalism differs from approaches existing in the literature
in that no superspace or superfield techniques are used.
Our derivation relies solely on algebraic techniques.
Supersymmetry originates from a superalgebra structure
underlying the integrable models.

In models based on the $sl(2|1)$ algebra 
the supersymmetry is associated with the symmetry generators
of the half-integer gradations. Our convention is such that the generator
$E$ of a space gradient has a unit gradation.
The supersymmetry transformations are realized as symmetry flows
generated by elements of half-integer grading in a 
centralizer $\cK$  associated with 
$E$. 
An anticommutator of elements of $\cK$ 
that square to $E$ 
generates a gradient. Thus, a fundamental property of 
supersymmetry is automatically satisfied.
The remaining integer generators in the centralizer $\cK$
give rise to non-local symmetries intertwining different
supersymmetry transformations.
Various $N=1$ and $N=2$ supersymmetric mKdV and sinh-Gordon equations
are derived from the $sl(2|1)$ algebra.
In contrast, the centralizer of the model based on the $osp(1|2)$ 
algebra does not contain any elements besides
the isospectral deformation generators. Thus, for this
model, there are  no symmetry flows apart from the isospectral times.

The Riemann-Hilbert problem naturally allows for an extension
to negative times \ct{Aratyn:2001pz}.
As an outcome, each positive integrable hierarchy 
is accompanied by a related negative hierarchy defined by the flow
equations of negative times.
As standard examples, the mKdV  and AKNS
hierarchies serve as positive hierarchies 
while the sine-Gordon and complex sine-Gordon hierarchies
serve as their negative counterparts \ct{dorfmeister,Aratyn:2000wr}.
Here, we extend the list of examples to the fermionic
extensions of the mKdV and  sinh-Gordon
hierarchies with or without supersymmetry.
We also show that the relativistic hierarchies, obtained by 
the reduction of the two-loop WZWN model \ct{Aratyn:1990tr},
can be identified with the flow of 
a negative hierarchy of an extended Riemann-Hilbert problem.

The content of the paper is as follows:
In Section \ref{section:ourapprach}, we extend the 
dressing-Riemann-Hilbert  procedure
to the negative times and the half-integer gradations.
We relate the negative hierarchy to contain the relativistic equations obtained
from the reduction of the WZNW model.
In this Section, we also construct  the pertinent zero-curvature 
equations and related conservation laws.
In addition to the family of
the conserved bosonic Hamiltonians we discover a new conserved
local fermionic charge.
Section \ref{section:mkdv} introduces the mKdV hierarchy, the sine-Gordon
hierarchy and its fermionic
extensions based on the $osp(1|2)$ algebra endowed with the principal gradation.
In Section \ref{section:sl21}, we construct the $N=2$ mKdV hierarchy
and the negative $N=2$ sinh-Gordon hierarchy for the $sl(2|1)$ algebra.
The $N=1$ mKdV hierarchy is obtained in 
Section \ref{section:reduction} via reduction employing 
the twisted automorphism. 
The reduced subalgebra is isomorphic to the $osp(2 |2)^{(2)}$ twisted 
supersymmetric loop algebra.

The supersymmetric formulation of the (modified) KdV 
and sinh-Gordon equations have been a subject of 
many papers in the last 10-15 years.
Here, we list a few of the contributions. They are based on 
an algebraic approach.

The supersymmetric KdV and mKdV equations
were introduced in  \ct{Mathieu:1987xz}, (see also
\ct{Kersten:ic}).  
Soon after, Mathieu and Labelle  
wrote down the equations for the $N=2$  supersymmetric KdV
equations \ct{Labelle:1990vv} (see also \ct{ziemek}). 
Kersten and Sorin  
studied the bihamiltonian structure of $N=2$ supersymmetric KdV model
within the superfield formalism \ct{Kersten:2002mr}.

Das and Galv\~{a}o 
have derived the supersymmetric KdV equation from 
the self-duality conditions imposed on four-dimensional 
Yang-Mills potential \ct{Das:vz}.
%with values in the $osp(2|1)$ algebra . 

In \ct{Delduc-Gallot}, Delduc and Gallot 
put forward a classification scheme of 
the supersymmetric integrable models in terms of
constant, odd integer grade generators of the loop
algebra that square to a semisimple element $E$.  
In a subsequent development, Miramontes and Madsen 
constructed supersymmetric non-local flows and 
corresponding non local conservation laws 
for a variety of models \ct{Madsen:1999ta} .

In \ct{Inami:1990ic,Inami:1990hk}, Inami and Kanno 
proposed a fermionic Lax associated to the $osp(2|2)^{(2)}$ algebra
and used it to obtain the supersymmetric mKdV and sinh-Gordon 
equations as well as their  conservation laws.
Their approach made use of the superfield formalism.
The $N=2$ KdV and  mKdV equations together with 
the generalized Miura transformation were formulated 
in \ct{Inami:1990hk} using the superspace notation and the Lie algebra $sl(2,2)^{(1)}$.
In this paper we use $sl(2,1)$ superalgebra which has a smaller rank.

\section{Our Approach}
\label{section:ourapprach}
\subsection{\sf The Riemann-Hilbert problem and the 
Dressing Formalism}
\label{subsection:rhproblem}
In an algebraic approach to integrable models 
the algebra of symmetries is identified with a centralizer 
of generators of the isospectral deformations \ct{Aratyn:2000sm}.
Recall, that the centralizer $\cK_\xi$ 
of an element $\xi$ in the algebra $\cG$ 
is the set of all elements commuting with $\xi$.
Due to Jacobi identity $\cK_\xi$ is a subalgebra.

We will be working with a loop algebra  $\cgh$ 
endowed with the principal gradation defined by a grading operator 
$Q$ to be given below for each specific model.
The gradation induces decomposition into graded
subspaces $ \cgh= \oplus_{n \in {\mathbb Z}} \, \cgh_n$ with  $\cgh_n$
such that $\lb Q \, , \, \cgh_n \rb = n\, \cgh_n$.

Furthermore, we define a  Gauss decomposition for a  group element
$g$ of the corresponding group $G$ which with respect to the given 
grading takes a form
\be
g= g_-\, g_+= N\, B\, M , \quad g_- =  N,\; g_+ =B\, M ,
\lab{gaussdec}
\ee
where 
\[
 N= \exp \( \cgh_{<} \), \;\;
M= \exp \( \cgh_{>} \), 
\]
are matrix exponentials constructed from  strictly negative  
$\cgh_{<} \subseteq \oplus_{n=-1}^{-\infty} \, \cgh_n$
and strictly positive
$\cgh_{>} \subseteq \oplus_{n=1}^{\infty} \, \cgh_n$
graded subalgebras.
$B$ is a group element of grade zero.

Another fundamental object in this setting
is a semisimple element $E^{(n)}$ of grade $n \in {\mathbb Z}$ 
which induces the decomposition ${\hat \cG}= \cK \oplus \cM$
where $\cK$  is a kernel and  $\cM$  is an 
image of the adjoint operation 
$ {\rm ad} (E^{(n)}) X = \left[ E^{(n)}, X \right]$.

The integrable structure we describe here is derived from
an extended Riemann-Hilbert factorization problem 
\ct{Aratyn:2001pz} :
\begin{equation}
\exp \(-{\sum_{n=1}^{\infty} E^{(n)} t_{n}}\) \,  
g \,
\exp \({\sum_{n=1}^{\infty} E^{(-n)} t_{-n}} \)
= \Theta^{-1} (t) \;  \Pi  (t) 
\lab{rh-def}
\end{equation}
where $g$ is a constant element in $G$ 
while the dressing matrix 
$\Theta = \exp \( \sum_{i<0} \theta^{(i)} \) $
is an exponential in $\cG_{<}$.
We further assume that the 
positive dressing matrix $\Pi$
decomposes as $ \Pi=B M$ where $B$ is non-singular matrix 
of grade zero while
$M = \exp \( \sum_{i>0} m^{(i)} \)$ is an exponential
power series expansion 
of elements in $\cG_{>}$.

By taking a derivative of identity \rf{rh-def}
with respect to $t_n$ and $ t_{-n}$ we obtain relations
\begin{alignat}{2}
\pder{}{t_n} \Theta (t)&= 
\(\Theta E^{(n)} \Theta^{-1}\)_{-} \Theta (t),
&\qquad \;\;\pder{}{t_n} \Pi (t) &=  -\(\Theta E^{(n)} \Theta^{-1}\)_{+}
 \Pi (t)
\lab{rheq}\\
\frac{\partial}{\partial t_{-n}} \Theta (t) 
&= - \left( \Pi E^{(-n)} \Pi^{-1} 
\right)_{-} \Theta (t), &\qquad\;\; \pder{}{t_{-n}} 
\Pi (t) &= \left( \Pi E^{(-n)} \Pi^{-1} \right)_{+} \Pi (t)
\lab{uthneg}
\end{alignat}
These equations define action of the isospectral flows 
$\pa / \pa t_n$ and the negative flows
$\pa / \pa t_{-n}$ on the dressing matrices.
In the above equations, $({\ldots} )_{+}$ denotes projection on
positive terms with grades $\geq 0$ in $\cgh$ and 
$({\ldots} )_{-}$ denotes projection on $\cgh_{-}$.

We now propose extension of equations \rf{rheq}-\rf{uthneg} 
to all positive grade
elements $K_i  $ in $\cK$  by associating to each such $K_i  $
a transformation $\d_{K_i}$ according to :
\br
\d_{K_i}\Theta &=& \(\Theta K_i \Theta^{-1}\)_{-} \Theta
\lab{kirheq}\\
\d_{K_i} \Pi
&=&  - \(\Theta K_i \Theta^{-1}\)_{+}
\Pi\, .
\lab{dkionm}
\er
The map $K_i \, \to \, \d_{K_i}$ is a homomorphism \ct{Aratyn:2000sm}:
\be
\left[ \d_{K_i} \,, \,  \d_{K_j} \right] \Theta =
\d_{\left[ {K_i} ,  {K_j} \right]} \Theta \, .
\lab{dhomo}
\ee
Applying, respectively $\pa^2 / \pa t_{\pm n} \pa t_{\pm k}$ 
and $\pa^2 / \pa t_{\pm k} \pa t_{\pm n} $ on both
sides of eq.\rf{rh-def} produces identical results due to commutativity
of $E^{( n)}$ with $E^{( k)}$. 
Commutativity of $K_i$ with $E^{( n)}$
ensures that the flows from \rf{kirheq} indeed define
symmetry transformations which commute with the isospectral flows 
$\pa / \pa t_{ n}$ .

These results can be extended to also yield commutativity 
of $\pa / \pa t_{ n}$ and $ \d_{K_i}$ 
with the negative flows $\pa / \pa t_{- k}$. 
For example taking a difference of
\[
\begin{split}
\d_{K_i} \pder{}{t_{-n}}  \Theta &= - \d_{K_i}  
\left( \Pi E^{(-n)} \Pi^{-1} 
\right)_{-} \Theta = \( \left[\(\Theta K_i \Theta^{-1}\)_{+} ,
\Pi E^{(-n)} \Pi^{-1} \right]\)_{-}  \Theta \\&- 
(\Pi E^{(-n)} \Pi^{-1})_{-} \(\Theta K_i \Theta^{-1}\)_{-}
 \Theta
\end{split}
\]
with
\[
\begin{split}
\pder{}{t_{-n}}  \d_{K_i}  \Theta &=\pder{}{t_{-n}} 
\(\Theta K_i \Theta^{-1}\)_{-} \Theta 
= -\( \left[\Pi E^{(-n)} \Pi^{-1} , 
\Theta K_i \Theta^{-1} \right]\)_{-}  \Theta \\&- 
 \(\Theta K_i \Theta^{-1}\)_{-}(\Pi E^{(-n)} \Pi^{-1})_{-}
 \Theta
\end{split}
\]
we obtain 
\[
\begin{split}
\( \d_{K_i} \pder{}{t_{-n}}- \pder{}{t_{-n}}  \d_{K_i} \) \Theta
\; \Theta^{-1} &= \left[ \(\Theta K_i \Theta^{-1}\)_{+},
\Pi E^{(-n)} \Pi^{-1}  \right]_{-}
+ \left[ \(\Theta K_i \Theta^{-1}\)_{-},
\(\Pi E^{(-n)} \Pi^{-1}\)_{-}  \right] \\
&+ \left[ \(\Pi E^{(-n)} \Pi^{-1}\)_{-},
\Theta K_i \Theta^{-1}\right]_{-} =0\, .
\end{split}
\]
Here the result is an identity
and follows without invoking the commutativity of the underlying 
algebra generators. It follows quite generally that the 
commutation relations involving flows 
generated by algebra elements of opposite grading 
vanish identically.

The structure of the Lax operator follows from the underlying 
Riemann-Hilbert problem through the following standard construction.
Let us take $n=1$ in eq. \rf{rheq} with $E \equiv E^{(1)}$ 
and identify $t_1$ with the space variable $x$.
Then :
\be
\begin{split}
\pa_x  (\Theta) &=\(\Theta E \Theta^{-1}\)_{-} \Theta
= \big\lbrack \Theta E \Theta^{-1} - \(\Theta E \Theta^{-1}\)_{+}
\big\rbrack \Theta  \\
&= \Theta  E - \( E+ \left[ \theta^{(-1)} , E \right] \) \Theta\\
&= \Theta  E - \( E+ A_0 \) \Theta
\end{split}
\lab{rht1sl2}
\ee
where $ A_0 = \left[ \theta^{(-1)} , E \right] $
is clearly in $\cM$ and of grade zero.
This leads to the dressing expression
\be
\Theta^{-1} \( \pa_x+ E+ A_0  \) \Theta
= \pa_x+ E
\lab{dreslaxsl2}
\ee
for the Lax operator $L =  \pa_x+ E+ A_0$. Similarly, for 
higher flows we obtain
\be
\Theta^{-1} \( \pder{}{t_n}+ E^{(n)}+ \sum_{i=0}^{n-1} D^{(i)}_n \) \Theta
= \pder{}{t_n}+ E^{(n)}
\lab{dresbn}
\ee
where
\[ \(\Theta E^{(n)} \Theta^{-1}\)_{+} =
E^{(n)}+ \sum_{i=0}^{n-1} D^{(i)}_n
\]
These dressing relations give rise to the zero-curvature conditions
\be
\left[ \pa_x+ E + A_0, \pder{}{t_n}+ E^{(n)}+ \sum_{i=0}^{n-1} D_n^{(i)}
\right]
= \Theta \left[ \pa_x+ E\, ,\, \pder{}{t_n}+ E^{(n)} \right]
\Theta^{-1 }  =0 \, .
\lab{z-curva}
\ee
The structure of the Lax operators
changes when terms with the half-integer grades 
appear in $\cgh = \oplus_{n \in {\mathbb Z}} \, \cgh_{n/2}$.
As a consequence of such terms 
being present in the exponent of the dressing matrix 
\[
\Theta = \exp \( \sum_{i<0} \theta^{(i)} \)
=
\exp \( \theta^{(-1/2)} +  \theta^{(-1)} +\theta^{(-\nicefrac{3}{2})} +{\ldots} \)
\, 
\]
the form of the Lax operator obtained by 
the dressing procedure is changed as follows.
\be
\begin{split}
\pder{}{t_1} (\Theta) &=\(\Theta E \Theta^{-1}\)_{-} \Theta
= \big\lbrack \Theta E \Theta^{-1} - \(\Theta E \Theta^{-1}\)_{+}
\big\rbrack \Theta  \\
&= \Theta  E + \( E+ \left[ \theta^{(-1)} , E \right] +
\left[ \theta^{(-1/2)} , E \right] + \h \left[ \theta^{(-1/2)} , \left[
\theta^{(-1/2)} , E \right]\right] \) \Theta \\
&= \Theta  E + \( E+ A_0 + A_{1/2} +k_0 \) \Theta
\end{split}
\lab{rht1}
\ee
here 
\br
A_0 &=& \left[ \theta^{(-1)} , E \right] + 
\h \left[ \theta^{(-1/2)} , \left[
\theta^{(-1/2)} , E \right]\right] \Big\v_{\cM}  \, \in \,\cM \lab{azero}\\
A_{1/2}&=& \left[ \theta^{(-1/2)} , E \right] \, \in \,\cM
\lab{aonehalf}\\
k_0 &=& \h \left[ \theta^{(-1/2)} , \left[
\theta^{(-1/2)} , E \right]\right] \Big\v_{\cK}
\, \in \,\cK
\lab{kzero}
\er
where $ \Big\v_{\cK}$ and $\Big\v_{\cM}$
denote projections on the kernel ${\cK}$ and image ${\cM}$, 
respectively.
This shows that, in case of the half-integer grading, the Lax 
operator should be defined as
\be 
\cL = \pa_x + E+ A_0 + A_{\nicefrac{1}{2}} +k_0 \,.
\lab{laxoper}
\ee
The unconventional grade zero term  $k_0$  which resides in $\cK$
is here present due to the  half-integer grading (encountered
in case of $sl(2|1)$ with principal gradation).

The dressing matrix $\Theta$ factorizes as
$ \Theta =U \, S$ \ct{Aratyn:2000sm} with $U$ which is entirely 
in $\cM$ and given by a local power series in components of
$A_0 $ and $ A_{\nicefrac{1}{2}}$. $U$ rotates the Lax operator 
$\cL$ into :
\be
U^{-1} \( \pa_x+ E+ A_0 + A_{1/2} +k_0 \) U
= \pa_x+ E +K^{(-)} \, , 
\lab{dreslaxu}
\ee
here the term 
$K^{(-)}= \sum_{i<0} k^{(i)}$
contains only  terms of a negative grade in $\cK$.
Additional factor $S$ in $ \Theta $ rotates
$\pa_x+ E +K^{(-)}$ into 
\be
S^{-1} \( \pa_x+ E +K^{(-)}\) S =
\pa_x + E, 
\lab{sgauge}
\ee 
by a non-local gauge transformation.

Consider now the case of a superalgebra 
with a kernel $\cK$ which contains 
a constant grade one-half element $D^{(\nicefrac{1}{2})}$.
According to \rf{kirheq} this term gives rise to the symmetry 
flow
\be
\pa_{\nicefrac{1}{2}} \Theta \equiv \d_{D^{(\nicefrac{1}{2})}}\Theta 
= \( \Theta D^{(\nicefrac{1}{2})} 
\Theta^{-1} \)_{-}
 \Theta \,.
\lab{defsusy}
\ee
We will show that $\pa_{\nicefrac{1}{2}}$-flow enters the zero-curvature
equation :
\be
 \left[ \pa_x +E +A_0 +A_{\nicefrac{1}{2}}+k_0 \, , \, \pa_{\nicefrac{1}{2}} + 
D^{(0)}+D^{(\nicefrac{1}{2})} \right]=0\, .
\lab{zchalf}
\ee
First, we rewrite the right hand side of eq. \rf{defsusy} as
\[
\begin{split}
\( \Theta D^{(\nicefrac{1}{2})} \Theta^{-1} \)_{-} \Theta &=
\left[ \Theta D^{(\nicefrac{1}{2})} \Theta^{-1} - 
\( \Theta D^{(\nicefrac{1}{2})} \Theta^{-1} \)_{+} \right] \Theta \\
&= \Theta D^{(\nicefrac{1}{2})} - D^{(\nicefrac{1}{2})}\Theta- \left[ \theta^{(-\nicefrac{1}{2})}, 
D^{(\nicefrac{1}{2})}  \right] \Theta
\end{split}
\]
Introducing
\[ D^{(0)} =  \left[ \theta^{(-\nicefrac{1}{2})}, 
D^{(\nicefrac{1}{2})}  \right] 
\]
we can rewrite the expression
\[
\pa_{\nicefrac{1}{2}} \Theta =  \Theta D^{(\nicefrac{1}{2})} - D^{(\nicefrac{1}{2})}\Theta- D^{(0)}
\Theta
\]
as a dressing formula
\be
\Theta \( \pa_{\nicefrac{1}{2}} +D^{(\nicefrac{1}{2})}\)  \Theta^{-1}=
\pa_{\nicefrac{1}{2}}+ D^{(0)} +D^{(\nicefrac{1}{2})} \, .
\lab{dressusy}
\ee
By dressing the obvious identity
\[
\left[\pa_x +E \, , \, \pa_{\nicefrac{1}{2}} +D^{(\nicefrac{1}{2})} \right] =0
\;\;\; \to \;\;\; 
\Theta \left[\pa_x +E \, ,\,  \pa_{\nicefrac{1}{2}} +D^{(\nicefrac{1}{2})} \right] \Theta^{-1}=
0
\]
we indeed arrive at the zero-curvature relation \rf{zchalf}.

Similarly, in presence of the half-integer grading
the zero-curvature condition \rf{z-curva} gives way to
\be
 \left[ \pa_x +E +A_0 +A_{\nicefrac{1}{2}}+k_0\, ,\, \pder{}{t_n} + 
E^{(n)}+\sum_{k=1}^n \(D_n^{(k)} +D_n^{(k-\nicefrac{1}{2})}\) 
\right]=0
\lab{zc1}
\ee

Let us now turn our attention to the negative 
flows described in eq. \rf{uthneg}.
Assuming decomposition
$\Pi=\Pi_0 \, \Pi_+  = B\,M$ with
$M = \exp \( \sum_{i>0} m^{(i)} \)$
we obtain
\be
\frac{\partial}{\partial t_{-n}} \Theta (t) 
= - \left(  B M  E^{(-n)} M^{-1} B^{-1}\right)_{-} 
\Theta (t)= -  B \left( M  E^{(-n)} M^{-1}\right)_{-}  B^{-1}
\Theta (t) \,.
\lab{uthnega}
\ee
Setting $n=1$ in equation \rf{uthnega} 
yields (for integer gradation)
\be
\frac{\partial}{\partial t_{-1}} \Theta 
%=  - \left(  B M E^{(-1)}M^{-1} B^{-1} \right)_{-} \Theta 
= - B  E^{(-1)} B^{-1}\, \Theta 
\lab{uth1eg}
\ee
In case of a system with the half-integer gradation 
the expression for $M$ becomes 
\[
M = e^{\cgh_{>}} = 
\exp \( m^{(\nicefrac{1}{2})} + m^{(1)} + m^{(\nicefrac{3}{2})}+{\ldots} \)
\]
and leads to an additional term in \rf{uth1eg}
\be
\begin{split}
\frac{\partial}{\partial t_{-1}} \Theta 
&=  - \left(  B \( E^{(-1)}+ \left[ E^{(-1)},  m^{(\nicefrac{1}{2})} \right]
\)B^{-1} \right)_{-} \Theta \\
&= - B  E^{(-1)} B^{-1} \Theta -
 B \jmath_{-\nicefrac{1}{2}}  B^{-1} \Theta , 
\end{split}
\lab{uth1feg}
\ee
where
\be
\jmath_{-\nicefrac{1}{2}}=\left[ E^{(-1)},  m^{(\nicefrac{1}{2})} \right]
\lab{jmonehaf}
\ee
We can rewrite relation \rf{uth1feg} as a dressing transformation:
\[
\Theta \frac{\partial}{\partial t_{-1}} \Theta^{-1} 
=\frac{\partial}{\partial t_{-1}} 
+ B \jmath_{-\nicefrac{1}{2}} B^{-1} +B E^{(-1)} B^{-1}
\]
Such analysis leads us to propose
the following three Lax operators :
\br
\cD_{+1}&=& \cL= 
\pa_x + A_0 +k_0+A_{\nicefrac{1}{2}}+E
%\pa_x -\pa_x B \,B^{-1}+ j_{+\nicefrac{1}{2}} +\cE_{+}
\lab{cdplus}\\
\cD_{+\nicefrac{1}{2}}&=& \pa_{\nicefrac{1}{2}}  + D^{(0)} +D^{(\nicefrac{1}{2})}
%\pa_{\nicefrac{1}{2}}  -\pa_{\nicefrac{1}{2}} B\, B^{-1} +D^{(\nicefrac{1}{2})}
\lab{cdhalf}\\
\cD_{-1}&=& \pa_{-1} + B \jmath_{-\nicefrac{1}{2}} B^{-1} +B E^{(-1)} B^{-1}
\lab{cdminus}
\er
where 
\[
\pa_{\nicefrac{1}{2}}, \quad \pa_{-1} = \pder{}{t_{-1}}
, \;\;\; {\rm and } \;\;\;\pa_x= \pder{}{t_{1}}
\]
are all commuting flows as follows from their definition
in \rf{rheq}, \rf{uthneg}, \rf{kirheq}, \rf{dkionm}. 
Applying the dressing procedure one easily sees
that this commutativity 
implies the zero-curvature equations :
\br
\left[ \cD_{+\nicefrac{1}{2}} \, , \, \cD_{-1} \right]&=&0
\lab{cdhalfcdminus}\\
\left[ \cD_{+\nicefrac{1}{2}} \, , \, \cD_{+1} \right]&=&0
\lab{cdhalfcdplus}\\
\left[ \cD_{+1} \, , \, \cD_{-1} \right]&=&0 \, .
\lab{cdpluscdminus}
\er
The brackets with $\cD_{+\nicefrac{1}{2}}$ act as compatibility equations which
define supersymmetry transformations and ensure invariance of
equations of motion (\rf{cdpluscdminus}), i.e.
\be
 \left[ \pa_x +E +A_0+k_0 +A_{\nicefrac{1}{2}}\, ,\,  \pa_{-1} + 
B \jmath_{-\nicefrac{1}{2}} B^{-1} +B E^{(-1)} B^{-1}
  \right]=0
\lab{zcneg}
\ee
under the supersymmetry transformation.
The argument goes as follows. For supersymmetry to preserve
\rf{cdpluscdminus} (or \rf{zcneg}) it must hold
that $ \pa_{\nicefrac{1}{2}} \left[ \cD_{+1} \, , \, \cD_{-1} \right]=0$.
This can be proved by using
\rf{cdhalfcdminus} and  \rf{cdhalfcdplus}
and  applying the Jacobi identity  to get
\be
\begin{split}
\pa_{\nicefrac{1}{2}} \left[ \cD_{+1} \, , \, \cD_{-1} \right]&=
- \left[ \left[ D^{(0)} +D^{(\nicefrac{1}{2})}, \cD_{+1} \right] , \cD_{-1} \right]
- \left[ \cD_{+1}, \left[ D^{(0)} +D^{(\nicefrac{1}{2})},  \cD_{-1} 
\right]  \right] \\
&=- \left[  D^{(0)} +D^{(\nicefrac{1}{2})},   \left[\cD_{+1}, \cD_{-1} 
\right]  \right] =0 \, .
\end{split}
\lab{jacobsym}
\ee
The grade $-1$ component of the zero-curvature relation \rf{zcneg}
\be
\pa_x \(B E^{(-1)} B^{-1}\) + 
 \left[ A_0 +k_0,\, B E^{(-1)} B^{-1}  \right]=0
\lab{grm1comp}
\ee
has a solution
\be
A_0 +k_0 = -\pa_{x} B\, B^{-1}\, .
\lab{idenaz}
\ee
The grade $-1$ component of equation 
\rf{cdhalfcdminus}
\[
\left[ \pa_{\nicefrac{1}{2}}  + D^{(0)} +D^{(\nicefrac{1}{2})}, \pa_{-1} + 
B \jmath_{-\nicefrac{1}{2}} B^{-1} +B E^{(-1)} B^{-1}
  \right]=0
\]
is equal to 
\[
\pa_{\nicefrac{1}{2}}\(B E^{(-1)} B^{-1}\)  +\left[ D^{(0)} ,
\, B E^{(-1)} B^{-1} \right] =0 \,.
\]
The obvious solution is
\be
D^{(0)}=  -\pa_{\nicefrac{1}{2}} B\, B^{-1} \, .
\lab{d0sol}
\ee
With solutions \rf{idenaz} and \rf{d0sol} we can rewrite $\cD_{+1}$ and
$\cD_{+\nicefrac{1}{2}} $ as
\br
\cD_{+1}&=& \cL= 
\pa_x -\pa_x B \,B^{-1}+ A_{\nicefrac{1}{2}} +E
\lab{cdplusa}\\
\cD_{+\nicefrac{1}{2}}&=& \pa_{\nicefrac{1}{2}}  -\pa_{\nicefrac{1}{2}} B\, B^{-1} +D^{(\nicefrac{1}{2})}
\lab{cdhalfa}
\er
We now write explicitly all the zero-curvature equations in 
components.
Equation \rf{cdhalfcdminus} implies
\br
\pa_{\nicefrac{1}{2}} \jmath_{-\nicefrac{1}{2}} &=& \left[ E^{(-1)} \, , \,  B^{-1} D^{(\nicefrac{1}{2})} B \right]
\lab{pahjmh}\\
\pa_{-1} \( \pa_{\nicefrac{1}{2}} B\, B^{-1} \)  &=& \left[
 B \jmath_{-\nicefrac{1}{2}} B^{-1} \, , \,  D^{(\nicefrac{1}{2})} \right] \, .
\lab{pampahbb}
\er
{}From \rf{cdhalfcdplus} we derive 
\br
\left[ E \, , \,  \pa_{\nicefrac{1}{2}} B\, B^{-1} \right] 
&=& 
\left[ A_{\nicefrac{1}{2}} \, , \,  D^{(\nicefrac{1}{2})} \right] 
 \lab{jhalfdhalf}\\
\pa_{\nicefrac{1}{2}}\, A_{\nicefrac{1}{2}} &=&  \left[\pa_{\nicefrac{1}{2}} B\, B^{-1}\, , \,  A_{\nicefrac{1}{2}} \right]
- \left[\pa_{x} B\, B^{-1}\, , \, D^{(\nicefrac{1}{2})} \right]\, .
\lab{pahjph}
\er
Finally, equation \rf{cdpluscdminus} yields
\br
\pa_{x}\, \jmath_{-\nicefrac{1}{2}} 
&=&  \left[E^{(-1)}\, , \,   B^{-1}  A_{\nicefrac{1}{2}}  B \right]
\lab{paxjmh}\\
\pa_{-1}\, A_{\nicefrac{1}{2}} &=&  \left[E\, , \,   B \jmath_{-\nicefrac{1}{2}}  B^{-1}  \right]
\lab{pamjph}\\
\pa_{-1} \( \pa_{x} B\, B^{-1} \)  &=& 
\left[B E^{(-1)} B^{-1}\, , \,E \right]
\, + \,\left[ B \jmath_{-\nicefrac{1}{2}} B^{-1}  \, , \,
A_{+\nicefrac{1}{2}} \right] \,.
\lab{pampaxbb}
\er
By dressing the identity
\[
\left[ \pa_{\nicefrac{1}{2}} +D^{(\nicefrac{1}{2})}
 \, , \, \pder{}{t_n} +E^{(n)}\right] =0
\;\;\; \to \;\;\; 
\Theta \left[  \pa_{\nicefrac{1}{2}} +D^{(\nicefrac{1}{2})}\, , \, \pder{}{t_n} +E^{(n)}
\right] \Theta^{-1}=
0
\]
we obtain the zero-curvature relation
\be
\left[ \pa_{\nicefrac{1}{2}}+D^{(0)} +D^{(\nicefrac{1}{2})}\, , \,
 \pder{}{t_n} + E^{(n)}+
 \sum_{k=1}^n \(D_n^{(k-1)} +D_n^{(k-\nicefrac{1}{2})}\) 
\right]=0
\lab{zc1h}
\ee
Denote for brevity 
\[
\cD_n = \pder{}{t_n} + E^{(n)}+
 \sum_{k=1}^n \(D_n^{(k-1)} +D_n^{(k-\nicefrac{1}{2})}\) \, , 
\]
then the above zero curvature condition
\[
 \left[ \cD_{\nicefrac{1}{2}} \, , \, \cD_n \right]=0
\]
ensures that the higher flows defined by
$\left[ \cD_{+1} \, , \, \cD_n \right]=0$ remain invariant
under the supersymmetry transformation.
Indeed, the statement 
$\pa_{\nicefrac{1}{2}} \left[ \cD_{+1} \, , \, \cD_n \right]=0$
follows from relations \rf{cdhalfcdplus} and \rf{zc1h}
by employing the Jacobi identity in the same way as it was done in
\rf{jacobsym}.

\subsection{\sf A Connection to the Relativistic Hierarchies}
\label{subsection:relativistic}
The negative hierarchy 
equations 
\rf{paxjmh}, \rf{pamjph}, \rf{pampaxbb}
can also be formulated by a Hamiltonian reduction procedure 
from the so-called two-loop WZNW model \ct{Aratyn:1990tr}. 
The WZNW model is defined in terms of 
two currents $J, {\bar J}$  of the loop algebra $\cgh$ :
\[
J = g^{-1} \pa g , \quad {\bar J} = {\bar \pa} g \, g^{-1}
\]
According to the Gauss decomposition \rf{gaussdec} 
we find 
\[
J = M^{-1} K M , \quad {\bar J} = N {\bar K} N^{-1} 
\]
where 
\br 
K &=& B^{-1} N^{-1}(\pa N) B + B^{-1} \pa  B +(\pa M ) M^{-1} 
\lab{Kdef}\\
\bar{ K} &=&  N^{-1}({\bar \pa} N) +  {\bar \pa}  B \, B^{-1}+
B {\bar \pa}  M \,  M^{-1}  B^{-1} 
\lab{bKdef}
\er
The WZNW currents satisfy the free equations of motion
\[
{\bar \pa} \( g^{-1} \pa g \) = 0, \quad
{\pa} \(  {\bar \pa} g\, g^{-1} \) = 0,
\]
which imply the following equations of motion 
\br
{\bar \pa} K + \left[ K, {\bar \pa}  M \,  M^{-1} \right]&=&0
\lab{keqmot}\\
{\pa} {\bar K} - \left[ {\bar K} , N^{-1}{ \pa} N  \right]&=&0
\lab{bkeqmot}
\er
Impose now the following constraints:
\br
J &=& E^{(-1)} + \jmath_{-\nicefrac{1}{2}} + \jmath_0 +  \jmath_{>} \lab{Jconst}\\
{\bar J}&=& E^{(1)} + {\bar \jmath}_{\nicefrac{1}{2}} + {\bar \jmath}_0 +  
{\bar \jmath}_{<} \lab{Jbconst}
\er
Note, that in the expressions for $J,{\bar J} $ we put to zero an 
infinite number of currents with grades $<-1$ and  $>1$, respectively
\ct{Aratyn:1990tr}. 

{}From expressions \rf{Kdef}-\rf{bKdef}
and the constraints \rf{Jconst}-\rf{Jbconst}
we find that
\br
B^{-1} N^{-1}(\pa N) B \big\v_{\rm constr} &=& E^{(-1)}  + \jmath_{-\nicefrac{1}{2}}
\lab{bnconst}\\
B {\bar \pa} M \, M^{-1} B^{-1} \big\v_{\rm constr} &=& 
E^{(1)}  + {\bar \jmath}_{\nicefrac{1}{2}}
\lab{pbnconst}
\er
Projecting equation of motion \rf{keqmot} on grades $-\nicefrac{1}{2}$ and
$0$ and using the constraint \rf{bnconst} we obtain :
\br
{\bar \pa} \jmath_{-\nicefrac{1}{2}} +\left[ E^{(-1)} \, ,\,  B^{-1} {\bar \jmath}_{\nicefrac{1}{2}}
B \right] &=& 0 \lab{jm12eq}\\
{\bar \pa} \( B^{-1} \pa  B \) +
\left[ E^{(-1)}  , B^{-1} E^{(1)}  B \right] 
+\left[ \jmath_{-\nicefrac{1}{2}} , B^{-1} {\bar \jmath}_{\nicefrac{1}{2}}  B \right] 
&=& 0 \lab{bm12eq}
\er
Projecting equation of motion \rf{bkeqmot} on grades $\nicefrac{1}{2}$ and
$0$ and using the constraint \rf{pbnconst} we obtain :
\br
{\pa} {\bar \jmath}_{\nicefrac{1}{2}} -\left[ E^{(1)} , B 
{\jmath}_{-\nicefrac{1}{2}} B^{-1} \right] &=& 0 \lab{bjm12eq}\\
{\pa} \( {\bar \pa}  B  B^{-1} \) -
\left[ E^{(1)}  , B E^{(-1)}   B^{-1} \right] 
-\left[  {\bar \jmath}_{\nicefrac{1}{2}}  , B \jmath_{-\nicefrac{1}{2}}  B^{-1} \right] 
&=& 0 \lab{bbm12eq}
\er
Notice, that conjugating equation \rf{bm12eq} with $B$ turns it 
into eq. \rf{bbm12eq}.
Thus both equations are equivalent.

Identifying 
\be
A_{\nicefrac{1}{2}}= {\bar \jmath}_{\nicefrac{1}{2}}, \;\;\; 
%D^{(-\nicefrac{1}{2})} = B \jmath_{-\nicefrac{1}{2}} B^{-1} , \;\;\; 
%D^{(-1)}= B E^{(-1)}   B^{-1}, \;\; \; 
A_0 = {\bar \pa} B\,  B^{-1}
\lab{identifya}
\ee
and
\be \pa_{-1} = \pa, \;\;\; \pa_x = - {\bar \pa}
\lab{identifyb}
\ee
turns eqs. \rf{jm12eq}, \rf{bjm12eq} and \rf{bbm12eq} into the 
zero curvature equations 
\rf{paxjmh}, \rf{pamjph} and \rf{pampaxbb}.
Below, we show explicitly that this identification holds for 
osp$(1|2)$ and sl$(2|1)$.

\subsection{\sf Conservation laws }
\label{subsection:conservation}
The main object here is a current density (see \ct{Aratyn:2000sm}):
\be 
\cJ_n = \Tr \(\left[ Q,\Theta \right] E^{(n)} \Theta^{-1}  \)
\lab{curr-dens}
\ee
Here $Q$ is the grading operator and the trace $\Tr$ includes projection on the zero grade
in addition to the usual matrix trace. 
Using that $\pa \Theta/\pa t_m  = \Theta E^{(n)}  - B_m \Theta$
as follows from \rf{rheq} for $
B_m =\( \Theta E^{(n)} \Theta^{-1}\)_{+} $
we find that the quantities 
\[
\pa_m \cJ_n = -  \Tr \(\left[ Q,B_m \right] \Theta E^{(n)} \Theta^{-1}  \)
, \qquad \pa_m = \pder{}{t_m}
\]
are local since $\Theta E^{(n)} \Theta^{-1}= U  E^{(n)} U^{-1}$.
Also, 
\[
\pa_m \cJ_n - \pa_n \cJ_m= \Tr \(\left[ Q,\Theta \right] 
\left[ E^{(m)}, E^{(n)}\right]  \Theta^{-1}  \)=0
\] 
Define now 
\[
\begin{split}
\cH_n=
-\pa_x \cJ_n &=   \Tr \((E +\h A_{\nicefrac{1}{2}})
\Theta E^{(n)} \Theta^{-1}  \) \\
&=
\Tr \(E \Theta E^{(n)} \Theta^{-1}  \)+\h \Tr \(A_{\nicefrac{1}{2}}
\Theta E^{(n)} \Theta^{-1}  \)\\
%&= - \cH_n
\end{split}
\]
where in obtaining the last equality we used equation \rf{rht1}.
%\[
%\pa_x (\Theta) = \pa_1 (\Theta)=  \Theta E - \( E+A_0 +k_0 +
%A_{\nicefrac{1}{2}}\) \Theta %= \Theta E - B_1 \Theta
%\]

It follows automatically that $\pa_m \cH_n - \pa_n \cH_m=0$.
Moreover, each $\int \cH_n$ defines a conserved charge
with respect to every isospectral flow as seen from
\[
\pa_m \int \cH_n =
- \int \pa_x \pa_m \cJ_n = - \pa_m \cJ_n \big\v^{\infty}_{-\infty} =0\, .
\]
The conclusion followed from locality
of  $\pa_m \cJ_n$. 
The lowest conserved density is
\[
\cH_1= \Tr \(E U E U^{-1}  \)+\h \Tr \(A_{\nicefrac{1}{2}}
U E U^{-1}  \)%= u^2-{\bar \psi} {\bar \psi}^{\pr}
\]
Consider now a quantity 
\be
\cF_n = \Tr \( D^{(\nicefrac{1}{2})} \Theta  E^{(n)} \Theta^{-1}  \)
\lab{cfn}
\ee
We will now derive the associated conservation laws.
We start with an expression
\[
\pa_n \cF_m=\pder{}{t_n}  
\Tr \( D^{(\nicefrac{1}{2})} \Theta  E^{(m)} \Theta^{-1}  \)=
\Tr \( D^{(\nicefrac{1}{2})} \left[
\( \Theta  E^{(n)} \Theta^{-1}  \)_{-} \, , \, 
 (\Theta  E^{(m)} \Theta^{-1})_{+} \right] \)
\]
from which it follows that
\[
\pa_m \cF_n - \pa_n \cF_m=0\, .
\]
Set from now $m=1$ in expression for 
$\pa_n \cF_m$. 
It simplifies to:
\be
\begin{split}
\pa_n \cF_1 &= \Tr \( D^{(\nicefrac{1}{2})} \left[
\( \Theta  E^{(n)} \Theta^{-1}  \)_{-} \, , \, 
 (\Theta  E \Theta^{-1})_{+} \right] \)\\ 
 &=
\Tr \( D^{(\nicefrac{1}{2})} \left[
\( \Theta  E^{(n)} \Theta^{-1}  \)_{-} \, , \, 
E - \left[E, \theta^{(-\nicefrac{1}{2})} \right] - \left[E,\theta^{(-1)} \right] \right.
\right.\\
&+ \left.\left.\h \left[ \left[E,  \theta^{(-\nicefrac{1}{2})} \right], \theta^{(-\nicefrac{1}{2})} \right]
\, \right] \)\\
& =\Tr \( D^{(\nicefrac{1}{2})} \left[
\( \Theta  E^{(n)} \Theta^{-1}  \)_{-} \, , \, 
E +A_{\nicefrac{1}{2}}+A_0 +k_0\, \right] \)
\end{split}
\lab{pancfm}
\ee
Adding the total derivative of a local quantity 
$-\Tr \( D^{(\nicefrac{1}{2})} \( \Theta  E^{(n)} \Theta^{-1}
\)_{-\nicefrac{1}{2}}\)$ to 
$\pa_n \cF_1$ in \rf{pancfm}
yields
\be
%\begin{split}
\pa_n \cF_1 - \pa_x \Tr \( D^{(\nicefrac{1}{2})} \( \Theta  E^{(n)} \Theta^{-1}
\)_{-\nicefrac{1}{2}}\)
%&= \Tr \( D^{(\nicefrac{1}{2})} \left[
%\( \Theta  E^{(n)} \Theta^{-1}  \)_{-} \, , \, \pa_x+
%E +A_{\nicefrac{1}{2}}+A_0+k_0 \, \right] \)\\
= \Tr \( D^{(\nicefrac{1}{2})} \left[
\( \Theta  E^{(n)} \Theta^{-1}  \)_{-} \, , \, \cL \right] \)
%\end{split}
\lab{panf1totder}
\ee
where $\cL=\pa_x+E +A_{\nicefrac{1}{2}}+A_0+k_0 $ is the Lax operator. 
Since $\cL$ is equal to
$\cL= \Theta (\pa_x +E) \Theta^{-1}$
it satisfies a commutation relation 
\[
\left[
\( \Theta  E^{(n)} \Theta^{-1}  \)_{-} \, , \, \cL \right]
= - \left[
\( \Theta  E^{(n)} \Theta^{-1}  \)_{+} \, , \, \cL \right]\, .
\]
The left hand side contains grades between $ \h$ and $-\infty$.
The right hand side contains positive ($\geq 0$) grades
and hence both sides must for consistency 
contain only grades $0$  and $\nicefrac{1}{2}$.
Therefore the right hand side of equation \rf{panf1totder}
vanishes after applying the trace and we arrive at:
\[
\pa_n \cF_1 = \pa_x \Tr \( D^{(\nicefrac{1}{2})} \( U  E^{(n)} U^{-1}
\)_{-\nicefrac{1}{2}} \)\, .
\]
Consequently, 
\be
\pa_n \int \cF_1 = 0, \qquad n > 0
\lab{cf1-cons}
\ee
expresses a new conservation law for the fermionic 
density $\cF_1 $.

\section{\sf The mKdV Hierarchy and its Extension to the
$osp(1|2)$ Algebra}
\label{section:mkdv}
In this section,  
we illustrate our construction for the case of 
the $sl(2)$ algebra with the principal gradation and show how to extend
the formalism to the $osp(1|2)$ algebra and include generators with 
the half-integer grading.

Consider first the $sl ( 2)$ algebra :
\[
\left[ H, E_{\pm} \right] = \pm 2 E_{\pm}, \quad
\left[ E_+, E_{-} \right] 
 = H\, .
\]
This structure generalizes straightforwardly to the loop algebra ${\hat \cG}$
spanned by the  generators $E_{\pm}^{(n)},
, H^{(n)}, n \in {\mathbb Z}$ (where $X^{(n)}=\l^n X$).
We define a principal gradation on ${\hat \cG}$
through the grading operator :
\be
Q = 2 d + \h H ,  \quad d = \l \dder{}{\l}
\lab{gradsl}
\ee
Grading of the generators $E_{\pm}^{(n)},
, H^{(n)}$ with respect to this grading operator $Q$ 
is defined by the commutation relations
\be 
\left[ Q , E_{\pm}^{(n)} \right] = (2n \pm 1) E_{\pm}^{(n)}
, , \;\; 
\left[ Q , H^{(n)} \right] = 2n H_{\pm}^{(n)}
\lab{gradesl2}
\ee
Define 
\be
E^{(2n+1)}=  E_+^{(n)}+ E_-^{(n+1)}
\lab{enpone}
\ee
The kernel and image of the adjoint operation 
$ {\rm ad} (E^{(2n+1)}) X = \left[ E^{(2n+1)}, X \right]$
are
\br
\cK &= & \{ E^{(2m+1)}\} =\{  E_+^{(m)}+ E_-^{(m+1)} \} 
\lab{kernelsl}\\
\cM &=& \{ E_+^{(m)}- E_-^{(m+1)} , H^{(m)} \} , \;\;
m \in {\mathbb Z} \,.
\lab{imosl}
\er
It is an important feature 
that $E^{(2n+1)}$ is semisimple 
and therefore induces relation ${\hat \cG}= \cK \oplus \cM$.

The semisimple element of grade one :
\be
E\equiv  E^{(1)}  = E_+^{(0)}+ E_-^{(1)}
\lab{sl2grad}
\ee
plays a particularly fundamental role in this formalism 
due to its connection to the Lax operator \ct{wilson}:
\be
\cL= \pa_x+ E + A_0=
\pa_x+ \begin{bmatrix}{0}&{1}\\
{\l}&0 \end{bmatrix} + \begin{bmatrix}v &{0}\\{0}&{-v}\end{bmatrix}\, , 
\lab{laxsl}
\ee
here $A_0$ is a grade zero element of $\cM$.

For $n=3$ in \rf{z-curva} we obtain:
\be
 \left[ \pa_x +E +A_0 , \pder{}{t_3} + 
D_3^{(0)}+D_3^{(1)} +D_3^{(2)}+ E^{(3)} 
 \right]=0 \,.
\lab{zcsl2}
\ee
Decomposing this relation according to grading 
one finds a solution
\[
D_3^{(0)}+D_3^{(1)} +D_3^{(2)}=
\frac{1}{4} \pa_x^2 A_0-\h v^3 H -\frac{1}{4} \left[ \pa_x A_0 \, , \,
E \right]
-\h v^2 E + \l^2 A_0\, .
\]
Projecting the zero-curvature equation \rf{zcsl2}
on its grade zero component yields 
$\pa A_{0}/\pa t_3 -\pa_x D_3^{(0)}- \left[ D_3^{(0)} , A_{0} \right]
=0$ which is equivalent to the celebrated mKdV equation:
\be
4 \pder{}{t_3} v = \pa_x^3 v - 6 v^2 \pa_x v \,.
\lab{mkdveq}
\ee
We will now show that 
a compatibility of the positive and negative 
flows expressed by a commutation relation 
\[
\left[ \cL , \cD_{-1} \right]
=\left[ 
\pa_x +E -\pa_x B \, B^{-1} , 
\pa_{-1} +B E^{(-1)} B^{-1}\right]=0
\]
%with $\cD_{+1} = \pa_{-1} +B E^{(-1)} B^{-1}$,
leads in this setting to the sinh-Gordon equation.
The zero grade component 
of $\left[ \cL , \cD_{-1} \right]=0$ is 
\be
\pa_{-1} \( \pa_{x} B\, B^{-1} \)  =
 \left[B E^{(-1)} B^{-1}\, , \,E \right]
\lab{sgordo}
\ee
Let the group element $B$ be given by
the exponential of the grade zero element of 
the algebra :
\be
B = e^{-\p H}
\lab{bph}
\ee
Plugging $B$ into equation  \rf{sgordo} results in
the sinh-Gordon equation
\be
\pa_x \pa_{-1} \p = e^{2 \p} - e^{-2 \p} \, .
\lab{sgordon}
\ee
We now generalize this formalism to the $osp (1 | 2)$ algebra. 
In the three-dimensional matrix representation its generators 
are
\[
H=\left[ 
{\begin{array}{cc|c}
1 & 0 & 0 \\
0 & -1 &  0 \tabularnewline\hline
0  & 0 & 0
\end{array}}
 \right] , \quad
E^{+}=\left[ 
{\begin{array}{cc|c}
0 & 1 & 0 \\
0 & 0 &  0 \tabularnewline\hline
0  & 0 & 0
\end{array}}
 \right] , \quad
E^{-}=\left[ 
{\begin{array}{cc|c}
0 & 0 & 0 \\
1 & 0 &  0 \tabularnewline\hline
0  & 0 & 0
\end{array}}
 \right]
\]
\[
F^+=\left[ 
{\begin{array}{cc|c}
0 & 0 & 1 \\
0 & 0 &  0 \tabularnewline\hline
0  & 1 & 0
\end{array}}
 \right] 
, \quad 
F^-=\left[ 
{\begin{array}{cc|c}
0 & 0 & 0 \\
0 & 0 & -1 \tabularnewline\hline
1  & 0 & 0
\end{array}}
 \right] \, .
\]

The corresponding loop algebra with generators $E_{\pm}^{(n)},
F_{\pm}^{(n)}, H^{(n)}, n \in {\mathbb Z}$ (where $X^{(n)}\
= \ \l^n X$)
reads 
\[
\left[ H^{(n)}, E_{\pm}^{(m)} \right] = \pm 2 E_{\pm}^{(n+m)}, \quad
\left[ H^{(n)}, F_{\pm}^{(m)} \right] = \pm  F_{\pm}^{(n+m)}, 
\]
\[
\left[ E_+^{(n)}, E_{-}^{(m)} \right] = 
\left\{ F_+^{(n)}, F_{-}^{(m)} \right\} = H^{(n+m)}
\]
\[
\left[ E_+^{(n)}, F_{-}^{(m)} \right] = - F_+^{(n+m)} , \;\;\;
\left[ E_-^{(n)}, F_{+}^{(m)} \right] = - F_-^{(n+m)} , 
%\;\;\;\left[ E_{\pm}, E_{\pm} \right] = 0
\]
\[
\left\{F_+^{(n)}, F_{+}^{(m)} \right\} = 2 E_+^{(n+m)} , \;\;\;
\left\{ F_-^{(n)}, F_{-}^{(m)} \right\} = - 2 E_-^{(n+m)}\, .
\]
The principal gradation 
is again defined through the grading operator \rf{gradsl}.
The novel feature is appearance of the half-integer 
grades as seen in  
\[ 
\left[ Q , E_{\pm}^{(n)} \right] = (2n \pm 1) E_{\pm}^{(n)}
, \;\; \left[ Q , F_{\pm}^{(n)} \right] = \(2n \pm \h \) F_{\pm}^{(n)}
, \;\; 
\left[ Q , H^{(n)} \right] = 2n H^{(n)}
\]
Consider again the semisimple element :
$E= E_+^{(0)}+ E_-^{(1)}\equiv  E^{(1)} $
of grade one. It defines the following kernel and image :
\br
\cK &= & \{ E^{(2n+1)}\} =\{  E_+^{(n)}+ E_-^{(n+1)} \} 
\lab{kernelosp}\\
\cM &=& \{ E_+^{(n)}- E_-^{(n+1)} , H^{(n)}, F_{\pm}^{(n)} \}
\lab{imosp}
\er
In the model under consideration the term $k_0$ in \rf{kzero} 
vanishes and the Lax operator is :
\be
L= \pa_x+ E +A_{\nicefrac{1}{2}} + A_0, \quad\;\; A_0 = v H^{(0)} , \;\;
A_{\nicefrac{1}{2}} = {\bar \psi}  F_{+}^{(0)}\, .
\lab{laxosp}
\ee
The zero curvature equation 
\be
 \left[ \pa_x +E +A_0 +A_{\nicefrac{1}{2}}, \pa_3 + 
D_3^{(0)}+D_3^{(\nicefrac{1}{2})} +D_3^{(1)} + D_3^{(\nicefrac{3}{2})} +D_3^{(2)}+ 
D_3^{(\nicefrac{5}{2})}+E^{(3)}
 \right]=0
\lab{zc2}
\ee
for the $t_3$-flow is solved by
\br
D_3^{(\nicefrac{5}{2})}&=& D^{(\nicefrac{5}{2})}_M = \l A_{\h}= {\bar \psi} F_+^{(1)} \nonu \\
D_3^{(2)}&=&D^{(2)}_M= \l A_{0} = v H^{(1)} \nonu \\
D_3^{(\nicefrac{3}{2})} &=& D^{(\nicefrac{3}{2})}_M=  \pa_x {\bar \psi} F_-^{(1)}
\nonu \\
D^{(1)}_{3\,M}&=&  \h \( {\bar \psi} \pa_x {\bar \psi}   + \pa_x v \) (E_+^{(0)} - E_-^{(1)})
 \nonu \\
D^{(1)}_{3\, K}&=& - \( \h v^2 + {\bar \psi}  \pa_x {\bar \psi} \)  (E_+^{(0)} + E_-^{(1)})
 \nonu \\
D_3^{(\nicefrac{1}{2})}&=&D^{(\nicefrac{1}{2})}_M= \( - v \pa_x {\bar \psi}  - \h {\bar \psi} \pa_x v +\pa^2_x {\bar \psi}
- \h {\bar \psi} v^2\) F_+^{(0)} \nonu \\
D_3^{(0)}&=& D^{(0)}_M= \frac{1}{4} \( \pa_x^2 v  - 2v^3 -6 v {\bar \psi} \pa_x {\bar \psi}+ 
3 {\bar \psi}  \pa_x^2 {\bar \psi} \) H^{(0)} \, .\nonu
\er
In this way we derive the equations of motion :
\br
4 \pa_3 v &=&\pa_x \( \pa_x^2 v- 2 v^3 -6 v  {\bar \psi} \pa_x {\bar \psi} +3 {\bar \psi}
\pa_x^2 {\bar \psi} \)
\nonu \\
&=& v^{\pr \pr \pr} - 6 v^2 v^{\pr } - 6 v^{\pr } {\bar \psi} {\bar \psi}^{\pr }
-6 v {\bar \psi} {\bar \psi}^{\pr \pr} + 3 {\bar \psi}^{\pr  } {\bar \psi}^{\pr \pr}
+ 3 {\bar \psi} {\bar \psi}^{\pr \pr\pr} \lab{vt3osp} \\
4 \pa_3 {\bar \psi} &=& 4 {\bar \psi}^{\pr \pr\pr}-6  v^{\pr }  {\bar \psi}^{\pr }
-6  v^{2 }  {\bar \psi}^{\pr }-6  v v^{\pr }  {\bar \psi}
- 3 v^{\pr \pr}  {\bar \psi} \, . \lab{xit3osp}
\er
A generalized version of the Miura transformation of the form :
\br
u&=& c \( v^{\pr } -  v^{2 }- {\bar \psi} {\bar \psi}^{\pr }\)   \lab{miuraa}\\
\eta &=& {\bar \psi}^{\pr }- v {\bar \psi} \lab{miurab} 
\er
(where $c$ is a non-zero constant parameter)
transforms the equations of motion \rf{vt3osp}-\rf{xit3osp}
into the fermionic extension of the KdV equation \ct{kuper}:
\br
4 \pa_3 u &=& u^{\pr \pr \pr} +\frac{6}{c} u u^{\pr} +12 c \,  
\eta \eta^{\pr \pr} \lab{kupera}\\
4 \pa_3 \eta &=&  4 \eta^{\pr \pr \pr} +\frac{6}{c} u \eta^{\pr} +
\frac{3}{c} u^{\pr} \eta  \lab{kuperb}
\er
Taking $c=-1$ in \rf{miuraa} and performing
 substitutions 
\[ t_3 \to t=t_3/4 , \;\; x \to -x , \;\;\; \eta \to \frac{\eta}{2}
\]
we arrive at
\br
\pa_t u &=& - u^{\pr \pr \pr} +6 u u^{\pr} -3\eta \eta^{\pr \pr} 
\lab{kuperaa}\\
\pa_t \eta &=&  -4 \eta^{\pr \pr \pr} +{6} u \eta^{\pr} +
{3} u^{\pr} \eta  \lab{kuperbb}
\er
which agrees with the Kupershmidt-KdV equation \ct{kuper,Mathieu:xy}

These equations are not invariant under supersymmetry transformation
in agreement with the kernel $\cK$ \rf{kernelosp}
possessing only generators of the isospectral deformations.

\subsection{\sf Dressing and conserved quantities for osp $(1 | 2)$ algebra}
\label{subsection:dressingosp}
We look for the lowest grade terms in 
$U^{-1} (\pa_x +E +A_0 +A_{\nicefrac{1}{2}}) U= \pa_x +E + \sum_{j=-\nicefrac{1}{2}}^{\infty}
k_j$.
On the level of grade $\h$ we get :
\[
A_{\nicefrac{1}{2}} + \left[ E ,  u^{(-\nicefrac{1}{2})} \right] = 0  \;\;
\to \;  u^{(-\nicefrac{1}{2})} =   {\bar \psi} F^{(0)}_-
\]
On the level of grade $0$ we find  :
\[
A_0 + \left[ E ,  u^{(-1)} \right] = 0  \;\;
\to \;  u^{(-1)} = - \h v \(E^{(0)}_{-} - E^{(-1)}_+ \)
\]
Contributions on both sides to grade $-\nicefrac{1}{2}$ are
\be
\pa_x u^{(-\nicefrac{1}{2})} + \left[ E ,  u^{(-\nicefrac{3}{2})} \right]+
\h \left[ A_0 ,  u^{(-\nicefrac{1}{2})} \right]+
\h \left[ A_{\nicefrac{1}{2}} ,  u^{(-1)} \right]= k_{-\nicefrac{1}{2}}
\lab{ospkminh}
\ee
Since the left  hand side
of \rf{ospkminh} is entirely in $\cM$ 
the term $k_{-\nicefrac{1}{2}}$ on the right hand side vanishes. 
Putting the left hand side
of \rf{ospkminh} to zero gives 
\[ u^{(-\nicefrac{3}{2})}= \(- \frac{3}{4} v {\bar \psi} + {\bar \psi}^{\pr} \) F^{(-1)}_-
\]
The expression for the first conserved Hamiltonian 
density gives now :
\[
\cH_1= \Tr \(E U E U^{-1}  \)+\h \Tr \(A_{\nicefrac{1}{2}}
U E U^{-1}  \)= -v^2-{\bar \psi} {\bar \psi}^{\pr}
\]
which satisfies
$\pa (-v^2-{\bar \psi} {\bar \psi}^{\pr})/ \pa t_3 = \pa_x ({\ldots} )$ for the 
Kupershmidt-mKdV flow from 
\rf{vt3osp}-\rf{xit3osp}.

In case of $osp (1|2)$ we define an analog of $\cF_n$ from 
\rf{cfn} as
\be
\cF_n = \Tr \( F_{+} \Theta  E^{(n)} \Theta^{-1}  \)
\lab{cfn-osp}
\ee
The argument leading to \rf{cf1-cons} also applies to 
osp $(1|2)$ and therefore we are interested in the conserved density:
\[
\begin{split}
\cF_1 &= \Tr \( F_{+} \Theta  E \Theta^{-1}  \)=
 - \Tr \( F_{+} \left[ E ,  u^{(-\nicefrac{3}{2})} \right]\)
+
\h \Tr \( F_{+} \left[\left[ E ,  u^{(-1)} \right],  
u^{(-\nicefrac{1}{2})} \right]\)\\
&+ 
\h \Tr \( F_{+} \left[\left[ E ,  u^{(-\nicefrac{1}{2})} \right],  u^{(-1)}
\right]\)
=
-\Tr \( F_{+} \left[ E ,  u^{(-\nicefrac{3}{2})} \right]\)
-\h \Tr \( F_{+} \left[ A_0 ,  u^{(-\nicefrac{1}{2})} \right]\)\\
&-\h \Tr \( F_{+} \left[ A_{\nicefrac{1}{2}} ,  u^{(-1)} \right]\)
= -\Tr \( F_{+} \pa_x u^{(-\nicefrac{1}{2})}\)
=- \pa_x{\bar \psi} 
\end{split}
\]
which clearly satisfies $\pa_n \cF_1= \pa_x( {\ldots} )$
in agreement with a general argument.

\subsection{\sf The Negative osp$(1|2)$ Hierarchy}
\label{subsection:negativeosp}
Let $B $ be given again  by $ e^{-\p H}$ as in \rf{bph}. This yields
$A_0= -\pa_x B \, B^{-1}  = \pa_x \p H$.
On the grade $ \pm \nicefrac{1}{2}$ level we set
\[
A_{\nicefrac{1}{2}} = {\bar \psi} F_+^{(0)}
= {\bar \psi}  F_+^{(0)}, \quad
\jmath_{-\nicefrac{1}{2}} = \psi F_-^{(0)}
\]
We now find that terms appearing in 
\rf{pamjph}, \rf{paxjmh} and  \rf{pampaxbb}
become in this parametrization
\br
B E^{(-1)} B^{-1}&=& e^{-2 \p} E_{+}^{(-1)} + e^{2 \p} E_{-}^{(0)}
\lab{dmin1}\\
B \jmath_{-\nicefrac{1}{2}} B^{-1}&=& \psi e^{\p} F_-^{(0)}\nonu\\
B^{-1} A_{\nicefrac{1}{2}} B&=& {\bar \psi} e^{\p} F_+^{(0)} \nonu 
\er
Equations \rf{pamjph}, \rf{paxjmh} and  \rf{pampaxbb}
correspond in this case to the following 
$osp(1|2)$ generalization of the sinh-Gordon equations
\br
\pa_{-1} {\bar \psi}+ \psi e^{\p}&=&0
\nonu\\
\pa_{x} \psi + {\bar \psi} e^{\p}&=&0
\lab{bbpp}\\
{\pa}_{-1} {\pa}_x \p - e^{2\p} + e^{-2 \p} 
+ \psi  {\bar \psi} e^{\p}&=&0
\nonu
\er

\section{\sf The $sl(2|1)$ Hierarchy; Unreduced $N=2$ Case}
\label{section:sl21}
\subsection{\sf $sl(2|1)$: $N=2$ mKdV from the zero curvature relations}
\label{subsection:sl21mkdv}
The starting point the 
Lax operator $\cL= \pa_x + A_0 +k_0+A_{\nicefrac{1}{2}}+E$
from \rf{cdplus} with $A_0$, $k_0$ and $A_{\nicefrac{1}{2}}$ being parametrized as
(see Appendix \ref{section:appendixa} for the matrix representation
of the $sl(2|1)$ algebra used here)
\be
A_0= u M_1^{(0)}+ v M_2^{(0)} 
= \left[ 
{\begin{array}{cc|r}
v&  - u {\displaystyle {\lambda }^{-1}}  & 0 \\ 
u \lambda  & -v & 0 \tabularnewline\hline
0 & 0 & 0
\end{array}}
\right] 
\lab{azpara}
\ee
\be
k_0 = \h \left[ \theta^{(-\nicefrac{1}{2})} , \left[
\theta^{(-\nicefrac{1}{2})} , E \right]\right] \Big\v_{\cK}
= {\bar \psi}_1  {\bar \psi}_2 \(K_2^{(0)}-K_1^{(0)}\)
\lab{kzerosl}
\ee
and 
\be
A_{\h} = {\bar \psi}_1 G_1^{(\h)} + {\bar \psi}_2 G_2^{(\h)}
=  \left[ 
{\begin{array}{cc|c}
0 & 0 & {\bar \psi}_1-{\bar \psi}_2 \\
0 & 0 & \lambda \({\bar \psi}_1-{\bar \psi}_2 \)
\tabularnewline\hline   
\lambda \Big({\bar \psi}_1+{\bar \psi}_2 \Big)  & {\bar \psi}_1+{\bar \psi}_2  & 0
\end{array}}
 \right]
\lab{ahpara}
\ee
where $ {\bar \psi}_i, i=1,2$ are anticommuting variables, i.e.
\[
{\bar \psi}_1^2 = {\bar \psi}_2^2 = 0 , \quad {\bar \psi}_1 {\bar \psi}_2 + {\bar \psi}_2 {\bar \psi}_1 = 0
\]
The detailed solution to  the zero curvature equation \rf{zc1}
with $n=3$ 
is given in Appendix \ref{section:appendixB}.
Solving this zero curvature equation leads to the following 
equations of motion 
\br
4 \pa_3 \bar \psi_1 &=&\pa_x \( \pa_x^2 \bar \psi_1- 2 (v^2-u^2)\bar \psi_1\)
+ 3 \bar \psi_2 \( v \pa_x u - u \pa_x v \) 
- u \( \bar \psi_1 \pa_x u-u \pa_x \bar \psi_1 \)\nonu \\
&+& v \( \bar \psi_1 \pa_x v- v \pa_x \bar \psi_1\)
\lab{xi1eqsmot}
\\
4 \pa_3 \bar \psi_2 &=&\pa_x \( \pa_x^2 \bar \psi_2- 2 (v^2-u^2)\bar \psi_2\)
+ 3 \bar \psi_1 \( v \pa_x u - u \pa_x v \) 
+ v \( \bar \psi_2 \pa_x v- v \pa_x \bar \psi_2\)\nonu \\
&-&u \( \bar \psi_2 \pa_x u-u \pa_x \bar \psi_2 \)
\lab{xi2eqsmot}\\
 4 \pa_3 u  &=& \pa_x 
\Bigl[  \pa_x^2 u - 2 (v^2-u^2) u -3 u \pa_x \bar \psi_1
\bar \psi_1+3 u \pa_x \bar \psi_2\bar \psi_2 -
3 v \pa_x (\bar \psi_1 \bar \psi_2) \Bigr]\nonu\\
&+& 4v \(  u \pa_x v- v \pa_x u \) +6v \( \pa_x \bar \psi_1 \pa_x \bar
\psi_2 +(v^2-u^2)\bar \psi_1 \bar \psi_2 \) 
\lab{ueqsmot}\\
4 \pa_3 v  &=& \pa_x \Bigl[  \pa_x^2 v - 2 (v^2-u^2) v -3 v \pa_x \bar \psi_1
\bar \psi_1+3 v \pa_x \bar \psi_2\bar \psi_2 
-3 u \pa_x (\bar \psi_1  \bar \psi_2) \Bigr]\nonu \\
 &+& 4u \(  u \pa_x v- v
\pa_x u \)
+6u\( \pa_x \bar \psi_1 \pa_x \bar \psi_2 + (v^2 -u^2)\bar \psi_1 \bar
\psi_2 \) \lab{veqsmot}
\er
These are the $N=2$ supersymmetric mKdV equations.

To find the supersymmetry transformations 
generated by a constant generator of supersymmetry of grade 
$\nicefrac{1}{2}$ :
\be
D^{(\nicefrac{1}{2})} = \eps_1 F_1^{(\nicefrac{1}{2})}+\eps_2 F_2^{(\nicefrac{1}{2})}
\lab{defdhalf}
\ee
we consider the zero-curvature equation \rf{zchalf}
with 
\be
D^{(0)} = M_1^{(0)} \( {{\bar \psi}}_1 \eps_1 - {{\bar \psi}}_2 \eps_2 \)
 + M_2^{(0)} \( {{\bar \psi}}_1 \eps_2 - {{\bar \psi}}_2 \eps_1 \) 
+ X_0 (K_2^{(0)}+K_1^{(0)}) 
\lab{dzerosuper}
\ee
where 
\be
X_0= \int \left[ u \( {\bar \psi}_1 \eps_2 - {\bar \psi}_2 \eps_1\)
+ v \( {\bar \psi}_2 \eps_2 - 
{\bar \psi}_1 \eps_1\) \right] \, .
\lab{xdef}
\ee
The zero-curvature equation \rf{zchalf} leads then
to the transformations of $u,v$:
\br
\pa_{\nicefrac{1}{2}} \, u &=& - \( {\bar \psi}_2^{\pr} \eps_2 -
{\bar \psi}_1^{\pr} \eps_1 \) + 2 v X_0 \lab{pahu}\\
\pa_{\nicefrac{1}{2}} \, v &=&  \( {\bar \psi}_1^{\pr} \eps_2 -
{\bar \psi}_2^{\pr} \eps_1 \) + 2 u X_0 \lab{pahv}
\er
and of the fermionic fields ${\bar \psi}_1, {\bar \psi}_2$ :
\br
\pa_{\nicefrac{1}{2}} \, {\bar \psi}_1 &=& u \eps_1-v \eps_2 - 2 {\bar \psi}_2
\, X_0  \lab{pahbps1}\\
\pa_{\nicefrac{1}{2}} \, {\bar \psi}_2 &=& u \eps_2-v \eps_1 - 2 {\bar \psi}_1
\, X_0 \, . \lab{pahbps2}
\er
The $N=2$ mKdV equations \rf{xi1eqsmot}, \rf{xi2eqsmot},
\rf{ueqsmot} and \rf{veqsmot} are invariant under the 
supersymmetry transformations 
\rf{pahu}, \rf{pahv}, \rf{pahbps1} and \rf{pahbps2}
as a result of $\cD_{\nicefrac{1}{2}}$ commuting with the 
Lax operators as in \rf{zc1h}
%\be
%\begin{split}
%&\left[ \pa_{\nicefrac{1}{2}} + 
%D^{(0)}+D^{(\nicefrac{1}{2})}, \pder{}{t_n}+ E^{(n)}+ \sum_{i=0}^{n-1} D_n^{(i)}
%\right]\\
%&= \Theta \left[  \pa_{\nicefrac{1}{2}} + 
%D^{(\nicefrac{1}{2})}\, ,\, \pder{}{t_n}+ E^{(n)} \right]
%\Theta^{-1 }  =0 , \quad n \geq 1 
%\end{split}
%\lab{z-curv12}
%\ee
and the usual dressing procedure arguments.

Associating $\d_{\pm}$ to $ \eps_{\pm} (F_1^{(\nicefrac{1}{2})} 
\pm F_2^{(\nicefrac{1}{2})} )$ 
according to \rf{kirheq} and using the homomorphism property
\rf{dhomo} we find the $N=2$ supersymmetry structure
\[
\left[ \d_{+} , \d_{-} \right] \Theta = 2 \eps_{+} \eps_{-} 
\d_E \Theta = 2  \eps_{+} \eps_{-} \pa_x \Theta ,
\qquad \left[ \d_{\pm} , \d_{\pm} \right] \Theta =0 \, .
\]

Note that the $N=2$ mKdV equations \rf{xi1eqsmot}-\rf{veqsmot}
reduce to the supersymmetric
mKdV systems \ct{Mathieu:1987xz,Mathieu:xy,Mathieu:talk} 
under reductions  :

\noindent
1) Set $v=0$ and ${\bar \psi}_2=0$. Then the flows for remaining
variables $u, {\bar \psi}={\bar \psi}_1$ are :
\br
4 \pa_3 u  &=& u^{\pr \pr\pr} +6 u^2 u^{\pr} + 3{\bar \psi} \(u {\bar \psi}^{\pr}\)^{\pr}
\lab{redu}\\
4 \pa_3 {\bar \psi}  &=& {\bar \psi}^{\pr \pr\pr}+3 u \(u {\bar \psi}\)^{\pr}
\lab{redxi1}
\er

\noindent
1) Set $u=0$ and ${\bar \psi}_1=0$. Then the flows for remaining
variables $v, {\bar \psi}={\bar \psi}_2$ are :
\br
4 \pa_3 v  &=& v^{\pr \pr\pr} -6 v^2 v^{\pr} - 3 {\bar \psi} \(v {\bar \psi}^{\pr}\)^{\pr}
\lab{redv}\\
4 \pa_3 {\bar \psi}  &=& {\bar \psi}^{\pr \pr\pr}-3 v \(v {\bar \psi}\)^{\pr}
\lab{redxi2}
\er
\subsection{\sf Relativistic sl$(2|1)$ Hierarchy with Supersymmetry, 
Unreduced Case}
\label{subsection:relaslunreds}
The starting point here is a zero-curvature equation \rf{zcneg} 
with $B$ such that it reproduces
\[
A_0+k_0 = u M_1^{(0)}+vM_2^{(0)} + {\bar \psi}_1 {\bar \psi}_2 (K_2^{(0)}-K_1^{(0)} )
\]
via relation
\[
A_0+k_0 ={\bar \pa} B \, B^{-1} \,.
\]
The following definition of $B$ 
\be
B = e^{\s K_1^{(0)}} e^{\rho K_2^{(0)}}
e^{\p M_1^{(0)}} e^{R M_2^{(0)}} 
\lab{bigb}
\ee
includes all generators in grade zero subalgebra.

For an arbitrary derivation $\d$, the vector
field $\d  B \, B^{-1} $ decomposes as 
\be
\begin{split}
\d  B \, B^{-1} &=  M_1^{(0)} \bigl( \d \p \cosh (2 \s) 
- \d R \cos (2 \p) \sinh (2 \s) \bigr) \\
&+ M_2^{(0)} \bigl(  \d R \cos (2 \p) \cosh (2 \s) 
-\d \p \sinh (2 \s) \bigr) \\
&+  K_2^{(0)}\d \rho + K_1^{(0)} \bigl( \d \s -  \d R \sin (2 \p) \bigr)
\end{split}
\lab{dbbinv}
\ee
Let 
\[ \jmath_{-\nicefrac{1}{2}} = \psi_1 G_1^{(-\nicefrac{1}{2})}+\psi_2 G_2^{(-\nicefrac{1}{2})}, \;\;\;\; 
%{\bar \jmath_{\nicefrac{1}{2}}} 
A_{\nicefrac{1}{2}}= {\bar \psi}_1 G_1^{(\nicefrac{1}{2})}+
{\bar \psi}_2 G_2^{(\nicefrac{1}{2})}\,.
\]
Calculating the $\cK$-component of \rf{pahjph} 
and using \rf{jhalfdhalf} to find the  $\cM$-component
of $\pa_{\nicefrac{1}{2}} B\, B^{-1} $
we obtain a constraint 
\footnote{Notice that substituting $j_{-\nicefrac{1}{2}}$ and 
$A_{\nicefrac{1}{2}}$ in 
the r.h.s of  eqn. \rf{pampaxbb}, it can be seen that it does not contain terms
proportional to $(K_1+K_2)$ and henceforth we can take consistently the
constraints \rf{vinculpaxbb} and \rf{cvinculpaxbb}.  Such type of constraints in the
$\cK$ subspace were considered in ref. \ct{dyonic} in order
to construct dyonic integrable models.}
\be
- \pa_{x} B\, B^{-1} \big\v_{\cK} = {\bar \psi}_1 {\bar \psi}_2\,
\( K_2^{(0)}-K_1^{(0)}\)
\lab{vinculpaxbb}
\ee
or in components using equation \rf{dbbinv} with $\d = \pa_{x}$
\be
\pa_{x} \rho = \pa_x R \, \sin (2 \p) - \pa_x \s
= - {\bar \psi}_1 {\bar \psi}_2
\lab{cvinculpaxbb}
\ee
Using equation  \rf{dbbinv}  we write  :
\be
-\pa_x  B \, B^{-1} =  M_1^{(0)} u  + M_2^{(0)} v 
+ {\bar \psi}_1 {\bar \psi}_2 \, ( K_2^{(0)}-K_1^{(0)})
\ee
where
\br
u &=&   \pa_x R \cos (2 \p) \sinh (2 \s) 
-\pa_x \p \cosh (2 \s)  \lab{uudef}\\
v &=&  \pa_x \p \sinh (2 \s)  
-\pa_x R \cos (2 \p) \cosh (2 \s) \lab{vvdef}
\er
We use eq. \rf{jhalfdhalf} together with
the identity
\be
\pa_{\nicefrac{1}{2}} \(\pa_{x} B\, B^{-1} \) - 
\pa_{x} \(\pa_{\nicefrac{1}{2}} B\, B^{-1} \) 
= \left[ \pa_{\nicefrac{1}{2}} B\, B^{-1} ,  \pa_{x} B\, B^{-1} \right]
\lab{pahpaxbb}
\ee
to obtain 
\be
%\begin{split} 
\pa_{\nicefrac{1}{2}} B\, B^{-1} = \( {\bar \psi}_2 \eps_2 - 
{\bar \psi}_1 \eps_1\)
M_1 - \( {\bar \psi}_1 \eps_2 - {\bar \psi}_2 \eps_1\)
M_2 
- E^{(0)} \,  X_0
%\end{split}
\lab{pahbbi}
\ee
Comparing equation \rf{pahbbi} and \rf{dbbinv} yields :
\br
 \pa_{\nicefrac{1}{2}} \p \cosh (2 \s) 
- \pa_{\nicefrac{1}{2}} R \cos (2 \p) \sinh (2 \s) 
&=&{\bar \psi}_2 \eps_2 -
{\bar \psi}_1 \eps_1  \lab{pahpra}\\
\pa_{\nicefrac{1}{2}} R \,  \cos (2 \p) \cosh (2 \s) 
-\pa_{\nicefrac{1}{2}} \p \sinh (2 \s) 
&=& {\bar \psi}_2 \eps_1 -{\bar \psi}_1 \eps_2 
\lab{pahprb}\\
\pa_{\nicefrac{1}{2}} \, \rho \,= \,\pa_{\nicefrac{1}{2}} \,\s - \pa_{\nicefrac{1}{2}} R \,\sin (2 \p)
&=& -X_0 \lab{pahrho}
\er
from which we derive :
\br
\pa_{\nicefrac{1}{2}} \, \p &=& \({\bar \psi}_2 \eps_2 -
{\bar \psi}_1 \eps_1 \) \cosh (2 \s) +
\({\bar \psi}_2 \eps_1 -{\bar \psi}_1 \eps_2 \)
\sinh (2 \s) \lab{pahpp}\\
\pa_{\nicefrac{1}{2}} R \cos (2 \p) &=& \({\bar \psi}_2 \eps_2 -
{\bar \psi}_1 \eps_1 \) \sinh (2 \s) +
\({\bar \psi}_2 \eps_1 -{\bar \psi}_1 \eps_2 \)
\cosh (2 \s) \lab{pahR}
\er
Note, that the transformations \rf{pahu}
\rf{pahv} also follow from the transformations 
\rf{pahpp}, \rf{pahR} through definitions \rf{uudef}-\rf{vvdef}.

{}From equations \rf{pahbps1}
\rf{pahbps2} we notice a simple identity :
\[
\pa_{\nicefrac{1}{2}} \, \( {\bar \psi}_1 {\bar \psi}_2 \) = 
\pa_x \, X_0
\]
from which it follows that 
\[
\pa_{\nicefrac{1}{2}} \, \rho = - X_0
\]
which confirms consistency of constraints \rf{cvinculpaxbb}
and \rf{pahrho}.

One can integrate the constraint \rf{cvinculpaxbb}
to derive expression for the auxiliary field $ \s$ 
in terms of dynamical fields  $ {\bar \psi}_1 ,
{\bar \psi}_2, R, \p$ as follows 
\be
\s = -\rho + \int \pa_x R \sin (2 \p)
=  \int \left[ {\bar \psi}_1 {\bar \psi}_2 +  \pa_x R\, \sin (2 \p) 
\right]
\lab{exprs}
\ee
from which 
\be
\pa_{\nicefrac{1}{2}} \, \s = - \pa_{\nicefrac{1}{2}} \,\rho + 
\pa_{\nicefrac{1}{2}} \int \pa_x R\, \sin (2 \p)
= X_0 + 
\pa_{\nicefrac{1}{2}} \int \pa_x R\, \sin (2 \p)
\lab{exprsh}
\ee
On the other hand we find from \rf{pahrho}
that 
\be
\pa_{\nicefrac{1}{2}} \, \s = \pa_{\nicefrac{1}{2}} \,\rho + \pa_{\nicefrac{1}{2}}R \,  \sin (2 \p) 
= - X_0+\pa_{\nicefrac{1}{2}}R \,  \sin (2 \p) 
\lab{exprsh2}
\ee
The identity 
\be
2 X_0 = \pa_{\nicefrac{1}{2}}R \,  \sin (2 \p) - 
\pa_{\nicefrac{1}{2}} \int \pa_x R\, \sin (2 \p)
\lab{idsh}
\ee
ensures compatibility of equation \rf{exprsh} with equation
\rf{exprsh2}
and shows that $\s$ is consistently defined via the integral in 
equation \rf{exprs}.

We collect below the equations of motion of the model
written in components :
\begin{alignat}{2}
\pa_{-1} {\bar \psi}_1 &= 
2 \( B \jmath_{-\nicefrac{1}{2}} B^{-1}\)_{G_2} ,
&\qquad \;\;
\pa_{-1} {\bar \psi}_2 &= 
2 \( B \jmath_{-\nicefrac{1}{2}} B^{-1}\)_{G_1}\\
\pa_{x} {\psi}_1 &= 2 \( B^{-1} A_{\nicefrac{1}{2}} B\)_{G_2},
&\qquad \;\;
\pa_{x} {\psi}_2 &= 2 \( B^{-1} A_{\nicefrac{1}{2}} B\)_{G_1}\\
\pa_{\nicefrac{1}{2}} {\psi}_1 &= 
2 \( B^{-1} D^{(\nicefrac{1}{2})} B\)_{G_2},
&\qquad \;\;
\pa_{\nicefrac{1}{2}} {\psi}_2 &= 2 \( B^{-1} D^{(\nicefrac{1}{2})} B\)_{G_1}
\end{alignat}
\vspace{-9mm}
\br
\pa_{-1} u &=& - 2 \cosh(2R) \sin(2\p)-2 \Bigl[
( B \jmath_{-\nicefrac{1}{2}} B^{-1})_{F_2}
{\bar \psi}_1-
( B \jmath_{-\nicefrac{1}{2}} B^{-1})_{F_1} {\bar \psi}_2 
\Bigr]
\\
\pa_{-1} v &=& - 2 \sinh(2R) -2 \Bigl[ ( B \jmath_{-\nicefrac{1}{2}} B^{-1})_{F_1}
{\bar \psi}_1-( B \jmath_{-\nicefrac{1}{2}} B^{-1})_{F_2} {\bar \psi}_2 
\Bigr]
\er
where $(B X B^{-1})_Y$ denotes a component of expression
$(B X B^{-1})$ along the $Y \in \cG$ direction.

\section{\sf Reduced Model and Supersymmetric KdV Equation}
\label{section:reduction}
\subsection{\sf Step Operators and Twisted Automorphism of $sl(2|1)$}
\label{subsection:twisted}
First, we define an automorphism of the finite dimensional algebra
through :
\br
\s (E_\a) &=& - E_{-\a} \lab{seapha}\\
\s (H) &=& - H \lab{shaut}
\er
where $E_\a$ are bosonic step operators
\[
E_\a=\left[ 
{\begin{array}{cc|c}
0 & 1 & 0 \\
0 & 0 &  0 \tabularnewline\hline
0  & 0 & 0
\end{array}}
 \right] \, .
\]
Recall, that roots of $sl(2|1)$ can be realized in terms 
of unit lengths vectors $e_i, f, \, i=1,2$ ($e_i \cdot e_j = \d_{ij},
f\cdot f =-1$) as:
\[ \a = e_1 -e_2,\,  \g= e_2-f, \, \a+\g = e_1-f\, .\]

The above automorphism is compatible with the 
bosonic commutation relations:
\[
\left[ H^i, E_\a \right] = \a^i E_\a  \; \to \;
\left[ \s(H^i), \s(E_\a) \right] = \a^i \s (E_\a )
\]
\[ \left[ E_\a, E_\b \right] = \eps (\a, \b) E_{\a+\b}  \; \to \;
\left[ \s(E_\a), \s (E_\b) \right] =  \eps (\a, \b) \s (E_{\a+\b})
\]
\[ \left[ E_\a, E_{-\a} \right] = \a . H  \; \to \;
\left[ \s(E_\a), \s (E_{-\a}) \right] =  \s (\a.H)
\]
The fermionic commutation relations :
\[ \{ F_\a , F_\b \} = \eps_F (\a, \b) E_{\a+\b}
\]
are consistent with the choice :
\be
\s (F_\a) = i F_{-\a}, \quad i = \sqrt{-1}
\lab{sfapha}
\ee
Here, the fermionic step operators are :
\[
F_\g=\left[ 
{\begin{array}{cc|c}
0 & 0 & 0 \\
0 & 0 &  1 \tabularnewline\hline
0  & 0 & 0
\end{array}}
 \right] 
, \qquad 
F_{\a+\g}=\left[ 
{\begin{array}{cc|c}
0 & 0 & 1 \\
0 & 0 &  0 \tabularnewline\hline
0  & 0 & 0
\end{array}}
 \right] \, .
\]
We now 
propose an extension of the automorphism $ \s \to {\hat \s}$ to the loop 
algebra of a form
\br
{\hat \s} \( E_\a^{(n)}\) &=& - (-1)^{\# ({E_\a^{(n)})}}
E_{-\a}^{(n+\eta_{E_\a})} \lab{hatsean}\\
{\hat \s} \( F_\g^{(n)}\) &=& i (-1)^{\#(F_\g^{(n)})}
F_{-\g}^{(n+\eta_{F_\g})} \lab{hatsfgn}
\er
where $\#(E_\a^{(n)})$ is the (principal) grade of $E_\a^{(n)}$ :
\[
\left[ Q, E_\a^{(n)}) \right] = \#(E_\a^{(n)}) \; E_\a^{(n)}\, .
\]
for the grading operator $ Q = \l \dder{}{\l} + \h \a.H$
from \rf{gradop}.
The term $(-1)^{\#(E_\a^{(n)})}$ corresponds to
the mapping $\l \to -\l$ used in the case of the automorphism 
for the homogeneous gradation \ct{vandeLeur:2000gk,Aratyn:2001pz}.
The numbers $ \eta_{E_\a}, \eta_{F_\g}$ are eigenvalues in 
\[
\left[ \a.H ,  E_\a \right] = \eta_{E_\a} E_\a 
, \qquad 
\left[ \a.H ,  F_\g \right] = \eta_{F_\g} F_\g \, .
\]

The bosonic generators of $\cM$ are of the form 
$M_2^{(n)} = \a.H^{(n)}$ and $ M_1^{(n)} = E_\a^{(n-1)}-
E_{-\a}^{(n+1)}$. From their behavior under the
 ${\hat \s}$ automorphism 
\[ 
{\hat \s} ( M_2^{(n)} ) = - (-1)^n M_2^{(n)} , \quad
{\hat \s} ( M_1^{(n)} ) =  (-1)^n  M_1^{(n)}
\]
we see that their zero grade components are respectively 
odd and even 
\be 
{\hat \s} ( M_2^{(0)} ) = -  M_2^{(0)} , \quad
{\hat \s} ( M_1^{(0)} ) =   M_1^{(0)}\, .
\lab{m12grade}
\ee
The fermionic generators of $\cM$ are:
\[
G_1^{(n+\h)} = \(F^{(n)}_{\a+\g}+F^{(n+1)}_{\g} \)+  \(
F^{(n+1)}_{-\a-\g}+F^{(n)}_{-\g} \)
\]
\[
G_2^{(n+\h)} = \(F^{(n)}_{\a+\g}+F^{(n+1)}_{\g} \)-  \(
F^{(n+1)}_{-\a-\g}+F^{(n)}_{-\g} \)
\]
and transform as :
\[ 
{\hat \s} ( G_2^{(n+\h)} ) =  (-1)^n G_2^{(n+\h)} , \quad
{\hat \s} ( G_1^{(n+\h)} ) = - (-1)^n  G_1^{(n+\h)}
\]
and so their $\h$-grade components are respectively, even and odd 
under the ${\hat \s}$  automorphism
\[ 
{\hat \s} ( G_2^{(\h)} ) =  G_2^{(\h)}, \quad
{\hat \s} ( G_1^{(\h)} ) = -  G_1^{(\h)}
\]
We also have 
\[ 
{\hat \s} ( K_2^{(n)} ) = - (-1)^n K_2^{(n)} , \quad
{\hat \s} ( K_1^{(n)} ) =   - (-1)^n K_1^{(n)}
\]
for
\[
K_1^{(n)} =  -E^{(n-1)}_{\a}+E^{(n+1)}_{-\a}
\]
and
\[
K_2^{(n)} =  \a.H^{(n)}+2 \g.H^{(n)}
\]
Also,
\[ 
{\hat \s} ( F_2^{(n+\h)} ) =  (-1)^n F_2^{(n+\h)} , \quad
{\hat \s} ( F_1^{(n+\h)} ) =   - (-1)^n F_1^{(n+\h)}
\]
for
the fermionic generators of $\cK$ :
\[
F_1^{(n+\h)} = \(F^{(n)}_{\a+\g}-F^{(n+1)}_{\g} \)+  \(
F^{(n+1)}_{-\a-\g}-F^{(n)}_{-\g} \)
\]
\[
F_2^{(n+\h)} = -\(F^{(n)}_{\a+\g}-F^{(n+1)}_{\g} \)+  \(
F^{(n+1)}_{-\a-\g}-F^{(n)}_{-\g} \)
\]

\subsection{\sf Reduced Model and its Symmetries }
\label{subsection:reduction}
To reduce the model we take equivalence classes under the automorphism
${\hat \s}$.
The even algebra elements define an invariant subalgebra under
${\hat \s}$ consisting of :
\[
\cM_{\rm Bose} = \{ M_1^{(2n)} = -E_\a^{(2n-1)} + E_{-\a}^{(2n+1)} 
, M_2^{(2n+1)} = \a.H^{(2n+1)} \}\] 
\[
\cM_{\rm Fermi} = \{ G_1^{(2n+\frac{3}{2})} 
 = \(F^{(2n+1)}_{\a+\g}+F^{(2n+2)}_{\g} \)+  \(
F^{(2n+2)}_{-\a-\g}+F^{(2n+1)}_{-\g} \)\, , 
\]\[ \, G_2^{(2n+\h)} = \(F^{(2n)}_{\a+\g}+F^{(2n+1)}_{\g} \)-  \(
F^{(2n+1)}_{-\a-\g}+F^{(2n)}_{-\g} \) \}
\]
The even elements in $\cK$ are :
\[
\cK_{\rm Bose} = \{ 
K_2^{(2n+1)} = \a.H^{(2n+1)} +2 \g .H^{(2n+1)}, \;
K_1^{(2n+1)} = - \( E_\a^{(2n)} +  E_{-\a}^{(2n+2)} \) 
 \}\] 
\[
\cK_{\rm Fermi} = \{ F_1^{(2n+\frac{3}{2})} 
 = -\(F^{(2n+1)}_{\a+\g}-F^{(2n+2)}_{\g} \)+  \(
F^{(2n+2)}_{-\a-\g}-F^{(2n+1)}_{-\g} \)\, ,
\]
\[ \, F_2^{(2n+\h)} = \(F^{(2n)}_{\a+\g}-F^{(2n+1)}_{\g} \)+ \(
F^{(2n+1)}_{-\a-\g}-F^{(2n)}_{-\g} \) \}
\]
We build the reduced model by projecting on even quantities
\[
X^{\rm red} = \h \( X + {\hat \s} (X) \)
\]
The even elements $X^{\rm red}$ constitute a subalgebra
isomorphic to the $osp(2 |2)^{(2)}$ twisted supersymmetric loop algebra.
It is interesting to note that the affine version of
this subalgebra appeared in
\ct{Toppan:1992qj} where it was used to
construct the $N=1$ supersymmetric conformal affine Toda model
based on the superfield formalism.
In \ct{Ivanov:aw} 
this construction was generalized to include
$N=2$ supersymmetric conformal affine Toda model
formulated in the superfield formalism based on 
$sl(2,2)^{(1)}$  supersymmetric KM algebra.

Note that $E$ is invariant under ${\hat \s}$. Thus
the reduced Lax operator is :
\be
\pa_x +E +A_0^{r} +A_{\nicefrac{1}{2}}^r +k_0^r
\lab{redlax}
\ee
with $A_0^r$ and $A_{\nicefrac{1}{2}}^r$ parametrized as
\[
A_0^r= u M_1, \quad A_{\nicefrac{1}{2}}^r = {\bar \psi} G_2^{(\h)}  
\]
and $k_0^r$ to be determined.

The third flow of $u,{\bar \psi}$ 
\br
4 \pa_3 u  &=& u^{\pr \pr\pr} +6 u^2 u^{\pr} - 3{\bar \psi} \(u {\bar \psi}^{\pr}\)^{\pr}
\lab{redua}\\
4 \pa_3 {\bar \psi}  &=& {\bar \psi}^{\pr \pr\pr}+3 u \(u {\bar \psi}\)^{\pr}
\lab{redxi1a}
\er
is obtained from \rf{xi1eqsmot}-\rf{veqsmot}
by setting $v=0,{\bar \psi}_1=0$.

Equations \rf{redua}-\rf{redxi1a} are the supersymmetric generalization
of mKdV equation (and can therefore be called super mKdV equations).
A super-Miura transformation :
\br
\vp &=& -i u^{\pr} +  u^2 - {\bar \psi} {\bar \psi}^{\pr} \lab{smiurap}
\\
\eta &=& -i {\bar \psi}^{\pr} + u {\bar \psi} \lab{smiuraeta}
\er
defines new fields $\vp, \eta$ which satisfy the 
the supersymmetric generalization
of KdV equation 
\br
4 \pa_3 \vp &=& \vp^{\pr \pr\pr} + 6  \vp \vp^{\pr}
-3 \eta \eta^{\pr\pr} \lab{skdvvp}\\
4 \pa_3 \eta &=& \eta^{\pr \pr\pr} +3 \( \eta \vp\)^{\pr}
\lab{skdveta}
\er

\subsection{\sf Dressing of the sl$(2,1)$ reduced model}
\label{subsection:dressingsl}
Dressing the reduced Lax operator \rf{redlax} according to
relation \rf{dreslaxu} and requiring that  the $\cM$-component
of $U^{-1} (\pa_x +E +A_0^{r} +A_{\nicefrac{1}{2}}^r +k_0^r) U$
vanishes i.e.:
\[
U^{-1} (\pa_x +E +A_0^{r} +A_{\nicefrac{1}{2}}^r +k_0^r) U \big\v_{\cM} =0
\]
provides a method to derive expressions for $u^{(-i)}, i=-\h, -1, -\frac{3}{2},{\ldots} $.
For the grade $0$ in $\cM$ we find :
\[
A_0^r + \left[ E ,  u^{(-1)} \right] = 0  \;\;
\to \;  u^{(-1)} = \frac{1}{4\l^2} \left[ A_0^r , E \right]
= \h u M_2^{(-1)}
\]
For the grade $\h$ in $\cM$ it holds that 
\[
A_{\nicefrac{1}{2}}^r + \left[ E ,  u^{(-\nicefrac{1}{2})} \right] = 0  \;\;
\to \;  u^{(-\nicefrac{1}{2})} = \frac{1}{4\l^2} \left[ A_{\nicefrac{1}{2}}^r  , E \right]
= -\h  {\bar \psi} G_1^{(-\nicefrac{1}{2})} \, .
\]
Calculating $k_0^r$ according to \rf{kzero} we find that 
it vanishes on the 
reduced manifold :
\[
k_0^r = \h \left[ u^{(-\nicefrac{1}{2})} , \left[
u^{(-\nicefrac{1}{2})} , E \right]\right]  =0\, .
\]

Contribution to $\cM$  of degree $-\nicefrac{1}{2}$ 
is found from the dressing formula to be:
\[
 \left[ E ,  u^{(-\nicefrac{3}{2})} \right]+
\h \left[\left[ A_{\nicefrac{1}{2}} , u^{(-\nicefrac{1}{2})} \right] , u^{(-\nicefrac{1}{2})} \right] +
\pa_x u^{(-\nicefrac{1}{2})}=0 ,\]
which yields
\[ u^{(-\nicefrac{3}{2})} =
\frac{1}{4\l^2} \left[\pa_x u^{(-\nicefrac{1}{2})}  , E \right]=
\frac{1}{16\l^3} \left[\left[ \pa_x A_{\nicefrac{1}{2}} , E\right], E \right]
= \frac{1}{4\l^2} \pa_x A_{\nicefrac{1}{2}}
= \frac{1}{4}  \pa_x {\bar \psi} G_2^{(-\nicefrac{3}{2})} \, .
\]
Contribution to $\cM$  of degree $-1$ 
is found in the reduced case  to be:
\[
 \left[ E ,  u^{(-2)} \right]+
\pa_x u^{(-1)}=0 ,\]
which gives
\[ u^{(-2)} =
\frac{1}{4 \l^2} \left[\pa_x u^{(-1)}  , E \right]=
\frac{1}{16\l^3} \left[\left[\pa_x A_{0} , E\right], E \right]
= \frac{1}{4 \l^2} \pa_x A_{0}
= \frac{1}{4}  \pa_x u M_1^{(-2)}\, .
\]
We next introduce $S=\exp(s)$ which enters equation \rf{sgauge}.
Contribution on $\cK$ of degree $-\nicefrac{1}{2}$ 
which are generated by the $U$ transformation are
\[
\h \left[ A_{\nicefrac{1}{2}} , u^{(-1)} \right] +
\h \left[ A_{0} , u^{(-\nicefrac{1}{2})} \right]=k_{-\nicefrac{1}{2}}=-\pa_x s^{(-\nicefrac{1}{2})}, \]
and can be gauged away 
according to \rf{sgauge} provided
\[
s^{(-\nicefrac{1}{2})} = - \h F_1^{(-\nicefrac{1}{2})} \int ({\bar \psi} u) \, .
\]
Similarly, from
\[
k_{-1}=\h\left[ A_0, u^{(-1)} \right] + 
\h \left[ A_{\nicefrac{1}{2}} , u^{(-\nicefrac{3}{2})}\right] 
= -\frac{1}{4}  {\bar \psi} {\bar \psi}^{\pr}\(K_2^{(-1)}- K_1^{(-1)}\) 
-\h u^2 K_1^{(-1)} 
\]
and $\pa s^{(-1)} + (\nicefrac{1}{2}) \left[ k_{-\nicefrac{1}{2}}, s^{(-\nicefrac{1}{2})} \right] 
+k_{-1}=0$ we obtain
\[
s^{(-1)} = \frac{1}{4\l^2} {\bar \psi} u \int ({\bar \psi} u ) E +
\h \int (u^2) K_1^{(-1)}+
\frac{1}{4}\(K_2^{(-1)}- K_1^{(-1)}\) 
\int ({\bar \psi} {\bar \psi}^{\pr} )\, .
\]
We also find
\[
k_{-\nicefrac{3}{2}} = \frac{3}{8} \( u^{\pr} {\bar \psi} - u {\bar
\psi}^{\pr} \) F_2^{(-\nicefrac{3}{2})} \, .
\]

\subsection{\sf Symmetries and Conservation Laws}
\label{subsection:symmetries}
According to the definition \rf{kirheq} of a symmetry transformation
we find that 
\br
\d_X A_{\nicefrac{1}{2}} &=& - \left[ E , \(\Theta X \Theta^{-1}\)_{-\nicefrac{1}{2}} \right]
\lab{dxahalf}\\
\d_X A_{0} &=& - \left[ E , \(\Theta X \Theta^{-1}\)_{-1} \right]
- \left[ A_{\nicefrac{1}{2}}, \(\Theta X \Theta^{-1}\)_{-\nicefrac{1}{2}} \Big\v_{\cK} \right]
\lab{dxazero}
\er
for $X \in \cK$.

We now set $X = \eps F_2^{(\nicefrac{1}{2})}$ with a constant Grassmannian 
parameter $\eps$ and calculate 
\[
\begin{split}
\(\Theta \eps F_2^{(\nicefrac{1}{2})} \Theta^{-1}\)_{-\nicefrac{1}{2}} &= 
\(US \eps F_2^{(\nicefrac{1}{2})} S^{-1}U^{-1}\)_{-\nicefrac{1}{2}} \\
&= \(U \( \eps F_2^{(\nicefrac{1}{2})}+\left[ s^{(-\nicefrac{1}{2})} , \eps F_2^{(\nicefrac{1}{2})}\right]
+\h \left[ s^{(-\nicefrac{1}{2})},
\left[ s^{(-\nicefrac{1}{2})}, \eps F_2^{(\nicefrac{1}{2})} \right]\right]\right.\right. \\
&+\left.\left.\left[ s^{(-1)}, \eps F_2^{(\nicefrac{1}{2})} \right] \)U^{-1}\)_{-\nicefrac{1}{2}}\\
&= \left[ u^{(-1)}, \eps F_2^{(\nicefrac{1}{2})} \right]
+ \left[ s^{(-1)}, \eps F_2^{(\nicefrac{1}{2})} \right]
\end{split}
\]
where we took into consideration that 
$\left[ s^{(-\nicefrac{1}{2})}, \eps F_2^{(\nicefrac{1}{2})} \right]=0$ 
and 
$\left[ u^{(-\nicefrac{1}{2})}, \left[ u^{(-\nicefrac{1}{2})}, \eps F_2^{(\nicefrac{1}{2})} \right]\right]=0$.
In this way we obtain :
\[
\(\Theta \eps F_2^{(\nicefrac{1}{2})} \Theta^{-1}\)_{-\nicefrac{1}{2}} =
-\frac{ \eps}{2} u  G_1^{(-\nicefrac{1}{2})} +\frac{\eps}{2} \int ( {\bar \psi}{\bar \psi}^{\pr}) F_1^{(-\nicefrac{1}{2})}
- \frac{ \eps}{2}\int (u^2)  F_1^{(-\nicefrac{1}{2})}
\]
Calculating, in the similar way 
\[
\(\Theta \eps F_2^{(\nicefrac{1}{2})} \Theta^{-1}\)_{-1} \big\v_{\cM} =
 \frac{ \eps}{2} {\bar \psi}^{\pr} M_2^{(-1)} - \frac{ \eps}{4}
\( \int (u^2)  - \int ( {\bar \psi}{\bar \psi}^{\pr}) \) 
{\bar \psi} M_2^{(-1)}
\]
we find that in \rf{dxahalf}-\rf{dxazero}
the non-local terms cancel and the expression for the supersymmetry
transformation \[\d_{\rm susy} = \d_{\eps F_2}\] becomes :
\be \d_{\rm susy} u=   \eps {\bar \psi}^{\pr} , \;\;\;\
 \d_{\rm susy} {\bar \psi} =   \eps u
\lab{susytrans}
\ee
These results also follow  by imposing the zero-curvature
condition \rf{zchalf} with $D^{(\nicefrac{1}{2})}={\eps F_2}$.

Correspondingly, the  supersymmetry
transformation of fields $\vp , \eta$ appearing in the supersymmetric
KdV equation \rf{skdvvp}-\rf{skdveta} is
\[
\d_{\rm susy} \vp=   \eps {\eta}^{\pr} , \;\;\;\
 \d_{\rm susy} {\eta} =   \eps \vp
\]
as follows from the super-Miura transformations 
\rf{smiurap}-\rf{smiuraeta}.

Note, that the element $K_2^{(1)}$ is an even element of the kernel
and therefore generates the symmetry transformation of the 
reduced hierarchy.
One can show that
\[
\begin{split}
\( \Theta K_2^{(1)} \Theta^{-1} \)_{-\nicefrac{1}{2}} &=
\( U S  K_2^{(1)} S^{-1}U^{-1} \)_{-\nicefrac{1}{2}}=
\( U (  K_2^{(1)} + \left[ s^{(-\nicefrac{1}{2})}, K_2^{(1)} \right]
+ \left[ s^{(-\nicefrac{3}{2})}, K_2^{(1)} \right] \right. \\
&+ \left. \h \left[ s^{(-\nicefrac{1}{2})}, 
\left[ s^{(-\nicefrac{1}{2})}, K_2^{(1)} \right] \right] )U^{-1} \)_{-\nicefrac{1}{2}}\\
&= \left[ u^{(-\nicefrac{3}{2})}, K_2^{(1)} \right]
+ \h \left[ u^{(-1)}, 
\left[ u^{(-\nicefrac{1}{2})}, K_2^{(1)} \right] \right]
+ \left[ u^{(-1)}, 
\left[ s^{(-\nicefrac{1}{2})}, K_2^{(1)} \right] \right]\\
&+ \left[ s^{(-\nicefrac{3}{2})}, K_2^{(1)} \right]+\h \left[ s^{(-1)}, 
\left[ s^{(-\nicefrac{1}{2})}, K_2^{(1)} \right] \right]
\end{split}
\]
Only the first and the third term on the right hand side
belong to $\cM$ and contribute to :
\[
\begin{split} 
\d_{K_2} A_{\nicefrac{1}{2}} &= - \Big\lbrack E \, , \, 
\( \Theta K_2^{(1)} \Theta^{-1} \)_{-\nicefrac{1}{2}} \Big\rbrack \\
&= -  \Big\lbrack E \, , \, \left[ u^{(-\nicefrac{3}{2})}, K_2^{(1)} \right]
+\left[ u^{(-1)}, 
\left[ s^{(-\nicefrac{1}{2})}, K_2^{(1)} \right] \right]
 \Big\rbrack
\end{split}
\]
A direct calculation gives :
\[
\left[ u^{(-\nicefrac{3}{2})}, K_2^{(1)} \right]
= \frac{1}{4} {\bar \psi}^{\pr} \left[ G_2^{(-\nicefrac{3}{2})} ,  K_2^{(1)} \right]=
- \frac{1}{4} {\bar \psi}^{\pr} G_1^{(-\nicefrac{1}{2})}
\]
\[
\left[ s^{(-\nicefrac{1}{2})}, K_2^{(1)} \right]
=- \h \int ({\bar \psi} u)  \left[ F_1^{(-\nicefrac{1}{2})} ,  K_2^{(1)} \right]=
\h \int ({\bar \psi} u) F_2^{(\nicefrac{1}{2})}
\]
and
\[
\left[ u^{(-1)}, 
\left[ s^{(-\nicefrac{1}{2})}, K_2^{(1)} \right] \right]
= \frac{1}{4} u \int ({\bar \psi} u) \left[ M_2^{(-1)} , F_2^{(\nicefrac{1}{2})}  \right]=
- \frac{1}{4} u \int ({\bar \psi} u) G_1^{(-\nicefrac{1}{2})}
\]
and so
\[\( \Theta K_2^{(1)} \Theta^{-1} \)_{-\nicefrac{1}{2}}\big\v_{\cM}
=-\frac{1}{4} \( u \int ({\bar \psi} u) + {\bar \psi}^{\pr} \) G_1^{(-\nicefrac{1}{2})} \, .
\]
Combining these results we get
\[
\d_{K_2} A_{\nicefrac{1}{2}} = - \Big\lbrack E \, , \, 
- \frac{1}{4} {\bar \psi}^{\pr} G_1^{(-\nicefrac{1}{2})}- 
\frac{1}{4}  u \int ({\bar \psi} u) G_1^{(-\nicefrac{1}{2})}
\Big\rbrack=
\h \({\bar \psi}^{\pr}  + u \int ({\bar \psi} u)\) G_2^{(\nicefrac{1}{2})}
\]
or
\[
\d_{K_2^{(1)}} {\bar \psi} = \h \({\bar \psi}^{\pr} + u \int ({\bar \psi} u) \) \, .
\]
Similarly, we find 
\[
\d_{K_2^{(1)}} u = - \h {\bar \psi}^{\pr}  \int ({\bar \psi} u) \,
\]
Using, that $\left[ K_2, F_2 \right]=F_1$
we obtain
\[
\left[ \d_{K_2^{(1)}}, \d_{\eps F_2^{(\nicefrac{1}{2})}} \right] {\bar \psi} = 
\d_{\eps F_1^{(\nicefrac{3}{2})}} {\bar \psi} = -\eps {\bar \psi}^{\pr} \int ({\bar \psi} u) 
-\h \eps u^{\pr} - \h \eps u \int (u)^2 + \h \eps u \int ({\bar \psi} {\bar \psi}^{\pr})
\]
\[
\left[ \d_{K_2^{(1)}}, \d_{\eps F_2^{(\nicefrac{1}{2})}} \right] u = 
\d_{\eps F_1^{(\nicefrac{3}{2})}} u = \h\eps {\bar \psi}^{\pr\pr} +\eps u^{\pr} \int ({\bar \psi} u) 
+\h \eps u^{2}{\bar \psi} - \h \eps {\bar \psi}^{\pr} \int (u)^2 
+ \h \eps {\bar \psi} \int ({\bar \psi} {\bar \psi}^{\pr})
\]
which agrees with transformation generated 
by $F_1^{\nicefrac{3}{2}}$.
One verifies that $\left[ \d_{\eps_1 F_1}, \d_{\eps_2 F_2} \right]
=0$ in agreement with $\left[ { F_1}, { F_2} \right]
=0$.
Hence repeated commutation relations with $ \d_{K_2}$ create the 
following chain 
\[ 
d_{\eps F_2^{(\nicefrac{1}{2})}}\, \stackrel{\d_{K_2^{(1)}}}{\longrightarrow} \, 
d_{\eps F_1^{(\nicefrac{3}{2})}} \,  \stackrel{\d_{K_2^{(1)}}}{\longrightarrow} \, 
d_{\eps F_2^{(\nicefrac{5}{2})}}\,  \stackrel{\d_{K_2^{(1)}}}{\longrightarrow} \, 
d_{\eps F_1^{(7/2)}} {\ldots} 
\]
of supersymmetry transformations of increasing grading.

Next, we turn to the discussion of conservation laws with respect to
the (bosonic) isospectral flows generated by 
$E^{(n)}$.
The main object here is a current density \rf{curr-dens}.
The lowest conserved density is
\[
\cH_1= \Tr \(E U E U^{-1}  \)+\h \Tr \(A_{\nicefrac{1}{2}}
U E U^{-1}  \)= u^2-{\bar \psi} {\bar \psi}^{\pr}
\]
where the fermionic contribution comes from
the term $\h \Tr \(A_{\nicefrac{1}{2}}
\Theta E^{(n)} \Theta^{-1}  \)$ present due to the half-integer 
gradation.

Another way to obtain $\cH_1$ is to take 
$\Tr \( E k_{-1}\) =-u^2+{\bar \psi} {\bar \psi}^{\pr}$
which must be conserved as a projection of $K^{-}$
on the $E$-direction, Note, that only $E$ is in the center of
$\cK$.

Applying the supersymmetry transformation on the dressing matrix we 
get:
\be
\begin{split}
\d_{\rm susy} \Theta &= \( \Theta \eps F_2^{(\nicefrac{1}{2})} \Theta^{-1} \)_{-}
 \Theta = \eps \Theta F_2^{(\nicefrac{1}{2})}- \( \Theta \eps F_2^{(\nicefrac{1}{2})} \Theta^{-1}
 \)_{+} \Theta \\
&= \eps \Theta F_2^{(\nicefrac{1}{2})} - \eps F_2^{(\nicefrac{1}{2})} \Theta  
+ \eps \cA_0 \Theta  , \quad  \cA_0 =
\left[  F_2^{(\nicefrac{1}{2})}, \theta^{(-\nicefrac{1}{2})} \right]
\end{split}
\lab{sus-theta}
\ee
Applying the above supersymmetry transformation on the
current density $\cJ_n$  from \rf{curr-dens} yields:
\be 
\d_{\rm susy} 
\cJ_n = -\h \eps \Tr \( F_2^{(\nicefrac{1}{2})} \Theta  E^{(n)} \Theta^{-1}  \)
\lab{susycurr-dens}
\ee
This leads to a Grassmannian quantity 
\be
\cF_n = \Tr \( F_2^{(\nicefrac{1}{2})} \Theta  E^{(n)} \Theta^{-1}  \)
\lab{cfnf2}
\ee
which is an $sl(2|1)$ analog of $\cF_n$ from 
\rf{cfn}.

Explicit calculation gives for $n=1$
\[
\cF_1 = {\bar \psi} u \, .
\]
In agreement with the general formalism this is a conserved
supercharge which satisfies $\pa ({\bar \psi} u )/ \pa t_3 = 
\pa_x ({\ldots} )$ for the $t_3$-mKdV flow from 
equations \rf{redua}-\rf{redxi1a}.

\subsection{\sf Relativistic sl$(2|1)$ Hierarchy in the Reduced Case
and its Supersymmetry Transformations}
\label{subsection:relativisticsl}
In this case,
\[ \jmath_{-\nicefrac{1}{2}} = \psi G_1^{(-\nicefrac{1}{2})}, \;\; 
{\bar \jmath_{\nicefrac{1}{2}}} = {\bar \psi} G_2^{(\nicefrac{1}{2})}, 
\;\;\;B = e^{\p M_1^{(0)}} \,.
\]
This parametrization leads to the following
expressions
\br
B^{-1} E^{(1)} B &=& K_2^{(1)} - \sin (2 \p)  M_2^{(1)}
+  \cos (2 \p)  K_1^{(1)} \lab{sl2bepb}\\
B E^{(-1)} B^{-1} &=& K_2^{(-1)} + \sin (2 \p)  M_2^{(-1)}
+  \cos (2 \p)  K_1^{(-1)} \lab{sl2bemb}\\
B^{-1}{\bar \jmath}_{\nicefrac{1}{2}} B &=& {\bar \psi} \(\cos (\p) G_2^{(\nicefrac{1}{2})}
+\sin (\p) F_2^{(\nicefrac{1}{2})} \) \lab{sl2bbjb}\\
B\jmath_{-\nicefrac{1}{2}} B^{-1} &=&{ \psi} \(\cos (\p) G_1^{(-\nicefrac{1}{2})}
-\sin (\p) F_1^{(-\nicefrac{1}{2})} \)\, . \lab{sl2bjbb}
\er
Plugging these results into eqs. \rf{jm12eq}, \rf{bjm12eq}, 
\rf{bm12eq} or \rf{bbm12eq} yields the relativistic 
supersymmetric sine-Gordon equations :
\be
{\pa}_x \psi =2 {\bar \psi} \cos ({\p}), \quad 
\pa_{-1} {\bar \psi} =2 \psi \cos  ({\p})
\lab{psibpsi-s-g}
\ee
\be
{\pa}_x {\pa}_{-1} \p =2 \sin ({2\p})
+ 2 \psi  {\bar \psi} \sin ({\p})
\lab{bbpp-s-g}
\ee
We find from equation \rf{jhalfdhalf} with 
$D^{(\nicefrac{1}{2})} = \eps F_2$
\be 
\pa_{\nicefrac{1}{2}} \p = \eps {\bar \psi}, 
\lab{susysingord}
\ee
Equation \rf{pahjmh} yields a supersymmetry transformation
\be
\pa_{\nicefrac{1}{2}} {\bar \psi}=
\eps \p^{\pr} \,.\lab{susysingorda}
\ee
This transformations agree with the supersymmetry transformations
\rf{susytrans} obtained directly from the dressing procedure
upon identification $ u=\p^{\pr}$.

Furthermore, equation \rf{pahjph} leads to
\be 
\pa_{\nicefrac{1}{2}} \psi = 2 \eps \sin \p
\lab{susysingordb}
\ee
The supersymmetry transformations \rf{susysingord}-\rf{susysingordb}
leave all the sine-Gordon equations \rf{psibpsi-s-g}-\rf{bbpp-s-g}
invariant.

\section{\sf Outlook}
\label{section:outlook}
This paper formulates supersymmetry for the integrable models 
based on superalgebras in an entirely algebraic fashion.
This framework provides a convenient set up 
for an application of the vertex operator construction to
deriving the soliton solutions \ct{Aratyn:1998ef}.
At first glance, it may appear surprising that the same vertex structure
would yield the soliton solutions of both the supersymmetric
sinh-Gordon and mKdV models. After all, the sinh-Gordon 
solitons are topological solitons while the mKdV solitons are 
hydrodynamic in nature.
Therefore, these two soliton configurations
are stable due to different reasons.
Note, however, that the vertex construction would be given in terms
of the gradient field of the mKdV equation : $u = \pa_x \p$.
The field $u$ is not affected by the kink-like behavior of $\p$ at 
$\pm \infty$. Hence, although both soliton solutions
would be  derived from the common algebraic structure 
the models remain defined in terms of different fields and some 
of their main characteristics disappear under a change of basis
($\p \to u$) performed when going from one model to another.

An extension of this paper will include other algebraic 
structures.
Work is being completed on the formalism 
of the supersymmetric AKNS and Lund-Regge
models defined for  superalgebras with the homogeneous gradation.
We also plan to extend our results to the higher rank algebras and 
describe supersymmetric
counterparts of the Boussinesq model.
\section*{\sf Acknowledgments}
H.A. acknowledges support from Fapesp and thanks the IFT Institute
for its hospitality. JFG and AHZ thank CNPq for a partial support.

\appendix
\section{\sf 
$sl(2|1)$ realization in principal gradation}
\label{section:appendixa}
The principal grading for the $sl(2|1)$
is generated by the operator
\be
Q= \l \dder{}{\l} + \h 
\left[\begin{array}{cr|c}
1&0&0 \\
0&-1&0\tabularnewline\hline
0&0&0\end{array}\right]
\lab{gradop}
\ee
Relation
\be
\left[ Q , E^{(n)}\right]=n E^{(n)} 
\lab{qefi} 
\ee
for 
\be
E^{(n)}= \l^{n-1} 
\left[\begin{array}{cr|c} \l & -1&0 \\
-\l^2&\l&0\tabularnewline\hline
0&0&2 \l 
\end{array}\right]
%\begin{bmatrix} \l & -1&0 \\
%-\l^2&\l&0\\
%0&0&2 \l 
%\end{bmatrix}
,
\lab{edefa}
\ee
shows that the semisimple element $ E=  E^{(1)} $ of $sl(2|1)$
has grade one.
It induces a decomposition of a loop algebra $Sl(2,1)$ in 
$\cM= Im (ad_E)$ and $\cK = Im (ad_E)$:
\[ Sl (2,1) = \cM \oplus \cK\]
The odd generators of $\cK$ are :
\be
F_1^{(n+\h)} = \l^n 
\left[\begin{array}{cr|r} 0 & 0 &1\\
0&0&-\l\tabularnewline\hline
\l&-1&0 
\end{array}\right]
%\begin{bmatrix} 0 & 0 &1\\
%0&0&-\l\\
%\l&-1&0 
%\end{bmatrix}
, \quad
F_2^{(n+\h)} = \l^n 
\left[\begin{array}{cr|r} 0 & 0 &-1\\
0&0&\l\tabularnewline\hline
\l&-1&0 
\end{array}\right]
%\begin{bmatrix} 0 & 0 &-1\\
%0&0&\l\\
%\l&-1&0 
%\end{bmatrix}
\lab{oddker}
\ee
The even generators of $\cK$ are :
\be
K_1^{(n)} = \l^n \left[ 
{\begin{array}{cc|r}
0 &  - {\displaystyle {\lambda }^{-1} }  & 0 \\ 
 - \lambda  & 0 & 0 \tabularnewline\hline
0 & 0 & 0
\end{array}}
 \right] , \quad
K_2^{(n)} = \l^n \left[ 
{\begin{array}{rr|r}
1 & 0 & 0 \\
0 & 1 & 0 \tabularnewline\hline
0 & 0 & 2
\end{array}}
 \right] 
\lab{evenker}
\ee
The odd generators of $\cM$ are :
\be
G_1^{(n+\h)} =\l^n  \left[ 
{\begin{array}{cr|c}
0 & 0 & 1 \\
0 & 0 & \lambda  \tabularnewline\hline
\lambda  & 1 & 0
\end{array}}
 \right] , \quad 
G_2^{(n+\h)} =\l^n  
\left[ 
{\begin{array}{cr|c}
0 & 0 & -1 \\
0 & 0 &  - \lambda \tabularnewline\hline
\lambda  & 1 & 0
\end{array}}
 \right] 
\lab{oddimage}
\ee
The even generators of $\cM$ are :
\be
M_1^{(n)} = \l^n\left[ 
{\begin{array}{cc|r}
0 &  - {\displaystyle {\lambda }^{-1}}  & 0 \\ 
\lambda  & 0 & 0 \tabularnewline\hline
0 & 0 & 0
\end{array}}
\right] , \quad
M_2^{(n)} = \l^n
 \left[ 
{\begin{array}{rr|r}
1 & 0 & 0 \\
0 & -1 & 0 \tabularnewline\hline
0 & 0 & 0
\end{array}}
 \right] 
\lab{evenimage}
\ee
Two odd generators
$F_i^{(\h)}, i=1,2$ have grade $\h$ according to 
\[
 \quad \left[ Q , F_i^{(n+\h)}\right]= {(n+\h)} F_i^{(n+\h)},\;\;
i=1,2
\]
The element $E$ is a square of $F_1^{(\h)}$ and
$F_2^{(\h)}$
\[
E= \(F_1^{(\h)}\)^2 = - \(F_2^{(\h)}\)^2
\]
which has a significance for the supersymmetry structure of the model.

The algebra can be split into $\left[ \cK \, , \, \cK \right] = \cK$
part containing :
\be
\left[ K_1^{(n)},  K_2^{(m)} \right] = 0
, \quad \{ F_1^{(n+\h)}, F_2^{(m+\h)} \} = 0, 
\lab{kwithk}
\ee
\be
\left[ F_1^{(n+\h)},  K_1^{(m)} \right] = F_2^{(n+m+\h)},
\;\;
\left[ F_1^{(n+\h)},  K_2^{(m)} \right] =- F_2^{(n+m+\h)},
\ee
\be
\left[ F_2^{(n+\h)},  K_1^{(m)} \right] = F_1^{(n+m+\h)},
\;\;
\left[ F_2^{(n+\h)},  K_2^{(m)} \right] =- F_1^{(n+m+\h)},
\ee
\be
  \{ F_1^{(n+\h)}, F_1^{(m+\h)} \} = 
-  \{ F_2^{(n+\h)}, F_2^{(m+\h)} \} = 2  E^{(n+m+1)}
\lab{ffalg}
\ee
the $\left[ \cK \, , \, \cM \right] = \cM$
part containing :
\be
 \{ F_2^{(n+\h)}, G_1^{(m+\h)} \}  
= -\{ F_1^{(n+\h)}, G_2^{(m+\h)} \}  
 = 2 M_1^{(n+m+1)}
\lab{fgalg1}
\ee
\be
 \{ F_1^{(n+\h)}, G_1^{(m+\h)} \}  
= -\{ F_2^{(n+\h)}, G_2^{(m+\h)} \}  
 = 2 M_2^{(n+m+1)}
\lab{fgalg2}
\ee
\be
\left[ M_1^{(n)},  F_1^{(m+\h)} \right] = G_1^{(n+m+\h)},
\;\;
\left[ M_1^{(n)},  F_2^{(m+\h)} \right] = G_2^{(n+m+\h)}
\ee
\be
\left[ M_2^{(n)},  F_1^{(m+\h)} \right] = -G_2^{(n+m+\h)},
\;\;
\left[ M_2^{(n)},  F_2^{(m+\h)} \right] = -G_1^{(n+m+\h)}
\ee
\be
\left[ M_1^{(n)},  K_1^{(m)} \right] = 2 M_2^{(n+m)},
\;\;
\left[ M_1^{(n)},  K_2^{(m)} \right] = 0
\ee
\be
\left[ M_2^{(n)},  K_1^{(m)} \right] = 2M_1^{(n+m)},
\;\;
\left[ M_2^{(n)},  K_2^{(m)} \right] = 0
\ee
\be
\left[ G_1^{(n+\h)},  K_1^{(m)} \right] = -G_2^{(n+m+\h)},
\;\;
\left[ G_1^{(n+\h)},  K_2^{(m)} \right] =- G_2^{(n+m+\h)},
\ee
\be
\left[ G_2^{(n+\h)},  K_1^{(m)} \right] = -G_1^{(n+m+\h)},
\;\;
\left[ G_2^{(n+\h)},  K_2^{(m)} \right] =- G_1^{(n+m+\h)},
\ee
and the part $\left[ \cM \, , \, \cM \right] = \cK$
characteristic for the symmetric space we have here
\be
\{ G_1^{(n+\h)}, G_2^{(m+\h)} \} = 0
\ee
\be
 \{ G_1^{(n+\h)}, G_1^{(m+\h)} \} = 
-  \{ G_2^{(n+\h)}, G_2^{(m+\h)} \} = 2 (K_2^{(n+m+1)}-
K_1^{(n+m+1)} )
\lab{ggalg}
\ee
\be
\left[ M_1^{(n)},  G_1^{(m+\h)} \right] = - F_1^{(n+m+\h)},
\;\;
\left[ M_1^{(n)},  G_2^{(m+\h)} \right] = -F_2^{(n+m+\h)}
\ee
\be
\left[ M_2^{(n)},  G_1^{(m+\h)} \right] = - F_2^{(n+m+\h)},
\;\;
\left[ M_2^{(n)},  G_2^{(m+\h)} \right] = -F_1^{(n+m+\h)}
\ee
\be
\left[ M_1^{(n)},  M_2^{(m)} \right] = -2 K_1^{(n+m)},
\ee
Some relations involving \[
E^{(n)} = K_1^{(n)}+K_2^{(n)}
\]
are
\be
\left[ M_1^{(n)},  E^{(m)} \right] = 2 M_2^{(n+m)},
\;\;
\left[ M_2^{(n)},  E^{(m)} \right] = 2 M_1^{(n+m)}
\ee
\be
\left[ G_1^{(n+\h)},  E^{(m)} \right] = -2 G_2^{(n+m+\h)},
\;\;
\left[ G_2^{(n+\h)},  E^{(m)} \right] = -2 G_1^{(n+m+\h)}
\ee
It follows  from them that
\be
\left[\left[ X_M^{(n)}, E^{(m)} \right] , E^{(p)} \right]
= 4 X_M^{(n+m+p)}
\lab{xnm}
\ee
for any $X_M \in \cM$.

Clearly, $\cgh/\cK$ is a symmetric space.

\section{\sf  Zero-Curvature Equations for $sl(2|1)$}
\label{section:appendixB}
Consider the zero-curvature equation \rf{zc1} with $n=3$ :
\be
 \left[ \pa_x +E +A_0 + A_{\nicefrac{1}{2}}+k_0, \pa_3 + 
D_3^{(0)}+D_3^{(\nicefrac{1}{2})} +D_3^{(1)} + D_3^{(\nicefrac{3}{2})} +D_3^{(2)}+ 
D_3^{(\nicefrac{5}{2})}+E^{(3)}
 \right]=0
\lab{zc3}
\ee
with $A_0$ , $k_0$ and $A_{1/2}$ parametrized as
$A_0= u_0 M_1^{(0)}+ v_0 M_2^{(0)} $ ,  
$k_0 = -\bar \psi_1  \bar \psi_2 (K_1^{(0)} -K_2^{(0)})$
and $A_{\h} = {\bar \psi}_1 G_1^{(\nicefrac{1}{2})} + 
{\bar \psi}_2 G_2^{(\nicefrac{1}{2})}$.
Below, for brevity we will suppres the subscript
${}_3$ in expressions for terms $D_3^{(i)}$.

The solution to the zero-curvature equation is 
obtained by decomposing the zero-curvature equation
according to grading into its $\cM= Im (ad_E)$
and $\cK= Ker (ad_E)$ components as follows.

Grade $7/2$ :
\be
%0&=&  \left[ D^{(3)}_M , A_{\nicefrac{1}{2}} \right] \lab{72k}\\
0= \left[  E , D^{(\nicefrac{5}{2})} \right]
+ \left[ A_{\nicefrac{1}{2}} ,  E^{(3)}  \right]
\lab{72m}
\ee
Grade $3$ :
\br
0&=&\pa_x D^{(3)}_K + \left[ A_{\nicefrac{1}{2}}, 
D^{(\nicefrac{5}{2})}_M    \right] 
% \left[ D^{(3)}_M , A_{0} \right] 
\lab{3k}\\
0&=& %\pa_x D^{(3)}_M+ 
\left[ E,  D^{(2)} \right]+ 
\left[ A_{\nicefrac{1}{2}},  D^{(\nicefrac{5}{2})}_K  \right]
+ \left[ A_{0}, E^{(3)}  \right] \lab{3m}
\er
Grade $\nicefrac{5}{2}$ :
\br
0&=&\pa_x D^{(\nicefrac{5}{2})}_K + 
\left[ A_{\nicefrac{1}{2}},  D^{(2)}_M   \right] +
 \left[ A_{0},  D^{(\nicefrac{5}{2})}_M \right] 
+ \left[ k_{0},  D^{(\nicefrac{5}{2})}_K  \right]
\lab{52k}\\
0&=& \pa_x D^{(\nicefrac{5}{2})}_M+ 
\left[  E , D^{(\nicefrac{3}{2})} \right]+ 
\left[ A_{\nicefrac{1}{2}},  D^{(2)}_K  \right]
+ \left[ A_{0},  D^{(\nicefrac{5}{2})}_K  \right]
+ \left[ k_{0},  D^{(\nicefrac{5}{2})}_M  \right]
\lab{52m}
\er
Grade $2$ :
\br
0&=&\pa_x D^{(2)}_K + 
\left[ A_{\nicefrac{1}{2}}, D^{(\nicefrac{3}{2})}_M   \right] +
 \left[ A_{0},  D^{(2)}_M \right] \lab{2k}\\
0&=& \pa_x D^{(2)}_M+ \left[  E , D^{(1)} \right]+ 
\left[ A_{\nicefrac{1}{2}}, D^{(\nicefrac{3}{2})}_K  \right]
+ \left[  A_{0} , D^{(2)}_K  \right] +
 \left[ k_{0},  D^{(2)}_M \right]
\lab{2m}
\er
Grade $\nicefrac{3}{2}$ :
\br
0&=&\pa_x D^{(\nicefrac{3}{2})}_K + 
\left[  A_{\nicefrac{1}{2}},  D^{(1)}_M  \right] +
 \left[  A_{0}, D^{(\nicefrac{3}{2})}_M  \right] +
 \left[  k_{0}, D^{(\nicefrac{3}{2})}_K  \right] 
\lab{32k}\\
0&=& \pa_x D^{(\nicefrac{3}{2})}_M+ 
\left[ E,  D^{(\nicefrac{1}{2})} \right]+ 
\left[ A_{\nicefrac{1}{2}}, D^{(1)}_K  \right]
+ \left[  A_{0}, D^{(\nicefrac{3}{2})}_K  \right] +
 \left[  k_{0}, D^{(\nicefrac{3}{2})}_M  \right] 
\lab{32m}
\er
Grade $1$ :
\br
0&=&\pa_x D^{(1)}_K + 
\left[ A_{\nicefrac{1}{2}},  D^{(\nicefrac{1}{2})}_M  \right] +
 \left[ A_{0}, D^{(1)}_M  \right] \lab{1k}\\
0&=& \pa_x D^{(1)}_M+ \left[ E , D^{(0)} \right]+ 
\left[  A_{\nicefrac{1}{2}}, D^{(\nicefrac{1}{2})}_K  \right]
+ \left[ A_{0} , D^{(1)}_K \right] +
 \left[ k_{0}, D^{(1)}_M  \right] 
\lab{1m}
\er
Grade $\nicefrac{1}{2}$ :
\br
0&=&\pa_x D^{(\nicefrac{1}{2})}_K + 
\left[ A_{\nicefrac{1}{2}},  D^{(0)}_M  \right] +
 \left[ A_{0} , D^{(\nicefrac{1}{2})}_M  \right]  +
 \left[ k_{0} , D^{(\nicefrac{1}{2})}_K  \right] 
 \lab{12k}\\
0&=& -\pa_3 A_{\nicefrac{1}{2}}+\pa_x D^{(\nicefrac{1}{2})}_M
+ \left[ A_{\nicefrac{1}{2}}, D^{(0)}_K   \right]
+ \left[ A_{0}, D^{(\nicefrac{1}{2})}_K  \right]
\lab{12m}
\er
Grade $0$ :
\br
0&=&\pa_x D^{(0)}_K -\pa_3 k_{0}+ \left[  A_{0}, D^{(0)}_M  \right] \lab{0k}\\
0&=& -\pa_3 A_{0}+\pa_x D^{(0)}_M+ \left[ A_{0}, D^{(0)}_K  \right] 
+ \left[ k_{0}, D^{(0)}_M  \right] \lab{0m}
\er
We found the following solution for the $D$'s :
\br
D^{(5/2)}&=& D^{(5/2)}_M = \l^2 A_{\h}\nonu \\
D^{(2)}_M&=& \l^2 A_{0} , \quad 
D^{(2)}_K = -\l^2 \bar \psi_1\bar \psi_2 \(K_1^{(0)}-K_2^{(0)}\) \nonu\\
D^{(3/2)}_M &=& -\h \l \( \pa_x \bar \psi_2\, G_1^{(\nicefrac{1}{2})} + 
\pa_x \bar \psi_1\, G_2^{(\nicefrac{1}{2})}\)
\nonu \\
D^{(3/2)}_k &=& - \h \l \( v \bar \psi_1 +u\bar \psi_2\) F_1^{(\nicefrac{1}{2})} 
- \h \l \( v \bar \psi_2 +u\bar \psi_1\) F_2^{(\nicefrac{1}{2})} \nonu \\
D^{(1)}_M&=&  \h \l \Big\lbrack \pa_x v M_1^{(0)}  + \pa_x uM_2^{(0)} - 
2 \bar \psi_1 \bar \psi_2
A_0 \Big\rbrack \nonu \\
D^{(1)}_K&=& -\h \l 
\( \pa_x \bar \psi_1 \bar \psi_1 -  \pa_x \bar \psi_2 \bar \psi_2 + ( v^2-u^2)
\) K_1^{(0)} +\h \l \( \pa_x \bar \psi_1 \bar \psi_1 -  \pa_x \bar \psi_2 \bar \psi_2 \) K_2^{(0)}
\nonu\\
D^{(1/2)}_M&=&  \( \frac{1}{4} \pa_x^2 - \h ( v^2-u^2) \)A_{\h}
\nonu \\
D^{(1/2)}_K&=& - \frac{1}{4} \( \pa_x u \bar \psi_1 - u \pa_x \bar \psi_1 +  
\pa_x v \bar \psi_2 - v \pa_x \bar \psi_2 \) F_1^{(\nicefrac{1}{2})}
\nonu\\ &-&
\frac{1}{4} \( \pa_x u \bar \psi_2 - u \pa_x \bar \psi_2 +  
\pa_x v \bar \psi_1 - v \pa_x \bar \psi_1 \) F_2^{(\nicefrac{1}{2})}
\nonu\\
D^{(0)}_M&=&  \frac{1}{4} \( \pa_x^2 u - 2(v^2-u^2)u- 
3 u \pa_x\bar \psi_1 \bar \psi_1
+ 3 u \pa_x\bar \psi_2 \bar \psi_2  -2  \bar \psi_1\bar \psi_2 \pa_x v- 
 v  \pa_x ( \bar \psi_1\bar \psi_2) \) M_1^{(0)} \nonu 
\\
&+&  \frac{1}{4} \( \pa_x^2 v - 2(v^2-u^2)v- 3 v \pa_x\bar \psi_1 \bar \psi_1 
+ 3v \pa_x\bar \psi_2 \bar \psi_2  -2 \bar \psi_1\bar \psi_2 \pa_x u - 
 u  \pa_x ( \bar \psi_1\bar \psi_2) \) M_2^{(0)} \nonu \\
D^{(0)}_K&=&  \h \( u \pa_x v - v \pa_x u+
 (v^2-u^2)\bar \psi_1\bar \psi_2 \)K_1^{(0)}\nonu\\
 &-& \frac{1}{ 4} \( \pa_x^2 \bar \psi_1 \bar \psi_2 + \bar \psi_1 \pa_x^2 \bar \psi_2 - \pa_x \bar \psi_1 \pa_x \bar \psi_2 - 3
 (v^2 - u^2) \bar \psi_1 \bar \psi_2 \) (K_1^{(0)} -K_2^{(0)})\nonu
\er


\begin{thebibliography}{99}
%\cite{Aratyn:2001pz}
\bibitem{Aratyn:2001pz}
H.~Aratyn, J.~F.~Gomes and A.~H.~Zimerman,
%``Integrable hierarchy for multidimensional Toda equations and  topological-anti-topological fusion,''
J.\ Geom.\ Phys.\  {\bf 46}, 21 (2003)
[Erratum-ibid.\  {\bf 46}, 201 (2003)]
[arXiv:hep-th/0107056].
%%CITATION = HEP-TH 0107056;%%
\bi{dorfmeister}
J. Dorfmeister, H. Gradl and J. Szmigielski, {\em Acta Applicandae Math.}
{\bf 53}, 1 (1998).
%\cite{Aratyn:2000wr}
\bibitem{Aratyn:2000wr}
H.~Aratyn, L.~A.~Ferreira, J.~F.~Gomes and A.~H.~Zimerman,
%``The complex Sine-Gordon equation as a symmetry flow of the AKNS
% Hierarchy,''
J.\ Phys.\  {\bf A 33}, L331 (2000)
[nlin.si/0007002].
%%CITATION = NLIN.SI 0007002;%%
%\cite{Aratyn:1990tr}
\bibitem{Aratyn:1990tr}
H.~Aratyn, L.~A.~Ferreira, J.~F.~Gomes and A.~H.~Zimerman,
%``Kac-Moody Construction Of Toda Type Field Theories,''
Phys.\ Lett.\ B {\bf 254}, 372 (1991).
%%CITATION = PHLTA,B254,372;%%
%\cite{Mathieu:1987xz}
\bibitem{Mathieu:1987xz}
P.~Mathieu,
%``Supersymmetric Extension Of The Korteweg-De Vries Equation,''
J.\ Math.\ Phys.\  {\bf 29}, 2499 (1988).
%%CITATION = JMAPA,29,2499;%%
%\cite{Kersten:ic}
\bibitem{Kersten:ic}
P.~H.~Kersten,
%``Higher Order Supersymmetries And Fermionic Conservation Laws Of The Supersymmetric Extension Of The Kdv Equation,''
Phys.\ Lett.\ A {\bf 134}, 25 (1988).
%%CITATION = PHLTA,A134,25;%%
%\cite{Labelle:1990vv}
\bibitem{Labelle:1990vv}
P.~Labelle and P.~Mathieu,
%``A New N=2 Supersymmetric Korteweg-De Vries Equation,''
J.\ Math.\ Phys.\  {\bf 32}, 923 (1991).
%%CITATION = JMAPA,32,923;%%
\bi{ziemek}
Z.~Popowicz,
%The Lax Formulation of the "New" N=2 SUSY KdV Equation,
Phys. Lett. {\bf A174}, 411 (1993).
%\cite{Kersten:2002mr}
\bibitem{Kersten:2002mr}
P.~H.~Kersten and A.~S.~Sorin,
%``Bi-Hamiltonian structure of the N=2 supersymmetric {\alpha}=1 KdV hierarchy,''
Phys.\ Lett.\ A {\bf 300}, 397 (2002)
[arXiv:nlin.si/0201061].
%%CITATION = NLIN-SI 0201061;%%
%\cite{Das:vz}
\bibitem{Das:vz}
A.~K.~Das and C.~A.~Galv\~{a}o,
%``Selfduality And The Supersymmetric KdV Hierarchy,''
Mod.\ Phys.\ Lett.\ A {\bf 8}, 1399 (1993)
[arXiv:hep-th/9211014].
%%CITATION = HEP-TH 9211014;%%
\bibitem{Delduc-Gallot}
F.~Delduc and L.~Gallot,
%Supersymmetric Drinfeld-Sokolov reduction
Jour. of Math. Phys. {\bf 39}, 4729 (1998) 
[arXiv:solv-int/9802013].
%\cite{Madsen:1999ta}
\bibitem{Madsen:1999ta}
J.~O.~Madsen and J.~L.~Miramontes,
%``Non-local conservation laws and flow equations for supersymmetric  integrable hierarchies,''
Commun.\ Math.\ Phys.\  {\bf 217}, 249 (2001)
[arXiv:hep-th/9905103].
%%CITATION = HEP-TH 9905103;%%
%\cite{Inami:1990ic}
\bibitem{Inami:1990ic}
T.~Inami and H.~Kanno,
%``Lie Superalgebraic Approach To Supertoda Lattice And Generalized Super Kdv Equations,''
Commun.\ Math.\ Phys.\  {\bf 136}, 519 (1991).
%%CITATION = CMPHA,136,519;%%
%\cite{Inami:1990hk}
\bibitem{Inami:1990hk}
T.~Inami and H.~Kanno,
%``N=2 Superkdv And Supersine-Gordon Equations Based On Lie Superalgebra A(1,1)**(1),''
Nucl.\ Phys.\ B {\bf 359}, 201 (1991).
%%CITATION = NUPHA,B359,201;%%
\bibitem{Aratyn:2000sm}
H.~Aratyn, J.~F.~Gomes, E.~Nissimov, S.~Pacheva and A.~H.~Zimerman,
%``Symmetry flows, conservation laws and dressing approach to the
% integrable models,''
in {\sl Integrable Hierarchies and Modern Physical Theories},
H. Aratyn and A. Sorin (eds.), Kluwer Academic Publ., pg. 243, (2001)
[nlin.si/0012042].
%%CITATION = NLIN.SI 0012042;%%
\bi{wilson}
G.~Wilson, Phil. Trans. R. Soc. London \ A {\bf 315}, 393 (1985).
\bi{kuper}
B.~Kupershmidt,  
%{\it A super Korteweg-de Vries equation: An integrable system}, 
Phys. Lett. {\bf A 102}, 213 (1984).
%\cite{Mathieu:xy}
\bibitem{Mathieu:xy}
P.~Mathieu,
%``Superconformal Algebra And Supersymmetric Korteweg-De Vries Equation,''
Phys.\ Lett.\ B {\bf 203}, 287 (1988).
%%CITATION = PHLTA,B203,287;%%
\bibitem{Mathieu:talk}
P.~Mathieu,
{\it Open problems for the superKdV equations},
in {\sl B\"{a}cklund and Darboux Transformations. 
The Geometry of Solitons},
Edited by: A. Coley, et al,
CRM Proceedings \& Lecture Notes, Volume: 29, 2001
[math-ph/0005007].
\bibitem{dyonic} J.F. Gomes, E.P. Gueuvoghlanian, G.M. Sotkov and A.H. Zimerman, 
Nucl. Phys. {\bf B598} (2001) 615-644,
also in hep-th/0011187;
Nucl.Phys. B {\bf 606}, 441  (2001) [hep-th/0007169].
%\cite{vandeLeur:2000gk}
\bibitem{vandeLeur:2000gk}
J.~van de Leur,
%``Twisted GL_n Loop Group Orbit and Solutions of the WDVV Equations,''
Intern. Math. Research Notices {\bf 11}, 551 (2001) %551- 574.
[nlin.si/0004021].
%%CITATION = NLIN.SI 0004021;%%
%\cite{Toppan:1992qj}
\bibitem{Toppan:1992qj}
F.~Toppan and Y.~Z.~Zhang,
%``Superconformal Affine Liouville theory,''
Phys.\ Lett.\ B {\bf 292}, 67 (1992)
[arXiv:hep-th/9208048].
%%CITATION = HEP-TH 9208048;%%
%\cite{Ivanov:aw}
\bibitem{Ivanov:aw}
E.~Ivanov and F.~Toppan,
%``N=2 Superconformal Affine Liouville Theory,''
Phys.\ Lett.\ B {\bf 309}, 289 (1993)
[arXiv:hep-th/9303073].
%%CITATION = HEP-TH 9303073;%%
%\cite{Aratyn:1998ef}
\bibitem{Aratyn:1998ef}
H.~Aratyn, L.~A.~Ferreira, J.~F.~Gomes and A.~H.~Zimerman,
%``Solitons from Dressing in an algebraic  approach to the constrained KP Hierarchy,''
J.\ Phys.\ A {\bf 31}, 9483 (1998) [solv-int/9709004];
{\sl Vertex Operators and  Solitons of
Constrained KP Hierarchies},in Lecture Notes in Physics 502
%{\it  Supersymmetry and Integrable Models},
%Proceedings of the UIC-Theory Workshop, June 1997,
H. Aratyn et al (Eds), Springer-Verlag, 1998, 
[solv-int/9711011].
\end{thebibliography}
\end{document}